\documentclass[iop]{emulateapj}

\usepackage{amsmath,amssymb,times,graphics,morefloats,color}
\usepackage{natbib,hyperref}
\newcommand{\kms}{~km~s\ensuremath{^{-1}}}
\newcommand{\masyr}{~mas~yr\ensuremath{^{-1}}}

\newcommand{\ra}{\ensuremath{\alpha}}
\newcommand{\dec}{\ensuremath{\delta}}
\newcommand{\pmra}{\ensuremath{\mu_{\alpha}}}
\newcommand{\pmdec}{\ensuremath{\mu_{\delta}}}
\newcommand{\dtheta}{\ensuremath{\Delta\theta}}
\newcommand{\dmu}{\ensuremath{\Delta\mu}}
\newcommand{\rz}{\ensuremath{r-z}}
\newcommand{\Pf}{P\ensuremath{_{\rm f}}}
\newcommand{\Mtot}{~M\ensuremath{_{\rm tot}}}
\newcommand{\Msun}{~M\ensuremath{_{\odot}}}

\newcommand{\slowpokes}{SLoWPoKES}
\defcitealias{Duquennoy1991}{DM91}
\defcitealias{Fischer1992}{FM92}
\defcitealias{Sesar2008}{SIJ08}
\defcitealias{Bochanski2010}{B10}
\shorttitle{\slowpokes}
\shortauthors{Dhital et al.}

\begin{document}
%\journalinfo{Submitted to AJ: December 7, 2009, Resubmitted: February 17, 2010}
\submitted{Submitted: December 7, 2009; Accepted: April 13, 2010}
\title{Sloan Low-mass Wide Pairs of Kinematically Equivalent Stars ({\slowpokes}):\\
  A Catalog of Very Wide, Low-mass Pairs}
\author{
  Saurav Dhital      \altaffilmark{1},
  Andrew A.\ West    \altaffilmark{2,3},
  Keivan G.\ Stassun \altaffilmark{1,4},
  John J.\ Bochanski \altaffilmark{3}}
\altaffiltext{1}{Department of Physics \& Astronomy, Vanderbilt University,
6301 Stevenson Center, Nashville, TN, 37235; saurav.dhital@vanderbilt.edu} 
\altaffiltext{2}{Department of Astronomy, Boston University, 725
  Commonwealth Avenue, Boston, MA 02215}
\altaffiltext{3}{Kavli Institute for Astrophysics and Space Research,
  Massachusetts Institute of Technology, Building 37, 77 Massachusetts Avenue, Cambridge, MA, 02139}
\altaffiltext{4}{Department of Physics, Fisk University, 1000 17th Ave.\ N.,
  Nashville, TN 37208}

\begin{abstract}
We present the Sloan Low-mass Wide Pairs of Kinematically Equivalent
Stars ({\slowpokes}), a catalog of 1342 very-wide (projected 
separation $\gtrsim 500$~AU), low-mass (at least one mid-K -- mid-M
dwarf component) common proper motion pairs identified from
astrometry, photometry, and proper motions in the  Sloan Digital Sky
Survey. A Monte Carlo based Galactic model  
is constructed to assess the probability of chance alignment for each
pair; only pairs with a probability of chance alignment $\leq 0.05$ are
included in the catalog. The overall fidelity of the catalog is
expected to be 98.35\%. The selection algorithm is purposely
exclusive to ensure that the resulting catalog is efficient for
follow-up studies of low-mass pairs. The {\slowpokes} catalog is 
the largest sample of wide, low-mass pairs to date and is
intended as an ongoing community resource for detailed study of {\em
  bona fide} systems. Here we summarize the general characteristics of
the {\slowpokes} sample and present preliminary results describing the
properties of wide, low-mass pairs.  While the majority of the
identified pairs are disk dwarfs, there are 70 halo subdwarf pairs and
21 white dwarf--disk dwarf pairs, as well as four triples. Most 
{\slowpokes} pairs violate the previously defined empirical limits for
maximum angular separation or binding energies. However, they are well
within the theoretical limits and should prove very useful in putting
firm constraints on the maximum size of binary systems and on
different formation scenarios. We find a lower limit to the wide
binary frequency for the mid-K -- mid-M spectral types that constitute
our sample to be 1.1\%. This frequency decreases as a function of
Galactic height, indicating a time evolution of the wide binary
frequency. In addition, the semi-major axes of the {\slowpokes} systems
exhibit a distinctly bimodal distribution, with a break at separations
around 0.1~pc that is also manifested in the system binding energy.
Comparing with theoretical predictions for the disruption of binary
systems with time, we conclude that the {\slowpokes} sample comprises
two populations of wide binaries: an ``old" population of tightly
bound systems, and a ``young" population of weakly bound systems that
will not survive more than a few Gyr.  The {\slowpokes} catalog and
future ancillary data are publicly available on the world wide web for
utilization by the astronomy community.
\end{abstract}

% ****************************************************
% ************       BODY          *******************
% ****************************************************

\section{Introduction}\label{Sec: intro}

The formation and evolution of binary stars remains one of the key
unanswered questions in stellar astronomy. As most stars are thought
to form in multiple systems, and with the possibility that binaries
may host exoplanet systems, these questions are of
even more importance. While accurate measurements of the fundamental
properties of binary systems provide constraints on evolutionary models
\citep[e.g.][]{Stassun2007}, knowing the binary frequency, as well as the
distribution of the periods, separations, mass ratios, and eccentricities of
a large ensemble of binary systems are critical to understanding binary
formation \citep[][and references therein]{Goodwin2007}. 
To date, multiplicity has been most extensively studied for the relatively 
bright high- and solar-mass local field populations \citep[e.g.][hereafter
\citetalias{Duquennoy1991}]{Duquennoy1991}. Similar studies of
low-mass M and L dwarfs have been limited by the lack of statistically
significant samples due to their intrinsic faintness. However, M
dwarfs constitute $\sim$70\% of Milky Way's stellar population
\citep[][hereafter \citetalias{Bochanski2010}]{Miller1979, Henry1999,
  Reid2002, Bochanski2010} and significantly influence its properties.

Since the pioneering study of \citet{Heintz1969}, binarity has been
observed to decrease as a function of mass: the fraction of primaries
with companions drops from 75\% for OB stars in clusters \citep{Gies1987,
  Mason1998, Mason2009} to $\sim$60\% for solar-mass stars
(\citealt{Abt1976}, \citetalias{Duquennoy1991},
\citealt{Halbwachs2003}) to $\sim$30--40\% for M dwarfs
(\citealt[][hereafter \citetalias{Fischer1992}]{Fischer1992}; \citealt{Henry1993,
  Reid1997, Delfosse2004}) to $\sim$15\% for brown dwarfs
\citep[BDs;][]{Bouy2003, Close2003, Gizis2003, Martin2003}. This
decrease in binarity with mass is probably a result of preferential
destruction of lower binding energy 
systems over time by dynamical interactions with other stars and
molecular clouds, rather than a true representation of the multiplicity
at birth \citep{Goodwin2005}. In addition to having a smaller total 
mass, lower-mass stars have longer main-sequence (MS) lifetimes
\citep*{Laughlin1997} and, as an ensemble, have lived longer and been
more affected by dynamical interactions. Hence, they are more
susceptible to disruption over their lifetime. Studies of young
stellar populations (e.g.\ in Taurus, Ophiucus, Chameleon) appear
to support this argument, as their multiplicity is twice as high as
that in the field \citep{Leinert1993, Ghez1997, Kohler1998}. However, 
in denser star-forming regions in the Orion Nebula Cluster and IC 348,
where more dynamical interactions are expected, the multiplicity is
comparable to the field (\citealt*{Simon1999}; \citealt{Petr1998};
\citealt*{Duchene1999b}). Hence, preferential destruction is likely to
play an important role in the evolution of binary systems.

\citetalias{Duquennoy1991} found that the physical separation of the
binaries could be described by a log-normal distribution, with the
peak at $a \sim$30~AU and $\sigma_{\log a} \sim$1.5 for F
and G dwarfs in the local neighborhood. The M dwarfs in the local
20-pc sample of \citetalias{Fischer1992} seem to follow a similar 
distribution with a peak at $a \sim$3--30~AU, a result severely 
limited by the small number of binaries in the sample. 

Importantly, both of these results suggest the existence of very
wide systems, separated in some cases by more than a parsec. 
Among the nearby ($d<$ 100~pc) solar-type stars in the {\em Hipparcos} catalog,
\citet*{Lepine2007a} found that 9.5\% have companions with projected
orbital separations $s>$ 1000~AU. However, we do not have
a firm handle on the widest binary that can be formed or on how they
are affected by localized Galactic potentials as they traverse the
Galaxy. Hence, a sample of wide binaries, especially one that spans a
large range of heliocentric distances, would help in (i) putting
empirical constraints on the widest binary systems in the field
\citep[e.g.][]{Reid2001b, Burgasser2003, Burgasser2007b,
Close2003, Close2007}, (ii) understanding the evolution of wide
binaries over time \citep*[e.g.,][]{Weinberg1987, Jiang2009},
and (iii) tracing the inhomogeneities in the Galactic potential
(e.g.~\citealt*{Bahcall1985} \citealt{Weinberg1987};
\citealt*{Yoo2004}; \citealt{Quinn2009}). 

Recent large scale surveys, such as the Sloan Digital Sky Survey
\citep[SDSS;][]{York2000}, the Two Micron All Sky Survey
\citep[2MASS;][]{Cutri2003}, and the UKIRT Infrared Deep Sky Survey
\citep[UKIDSS;][]{Lawrence2007}, have yielded samples of unprecedented
numbers of low-mass stars. SDSS alone has a photometric catalog of
more than 30 million low-mass dwarfs \citepalias{Bochanski2010},
defined as mid-K -- late-M dwarfs for the rest of the paper and a
spectroscopic catalog of more than 44000 M dwarfs
\citep{West2008}. The large astrometric and photometric catalogs of
low-mass stars afford us the opportunity to explore anew the binary
properties of the most numerous constituents of  Milky Way,
particularly at the very widest binary separations.

The orbital periods of very wide binaries (orbital separation $a>$ 100~
AU) are much longer than the human timescale ($P=$ 1000~yr for
\Mtot $=$ 1\Msun\ and $a=$ 100~AU). Thus, these systems can only be
identified astrometrically, accompanied by proper motion or radial
velocity matching. These also remain some of the most under-explored
low-mass systems. Without the benefit of retracing the binary orbit,
two methods have been historically used to identify very wide pairs:
\begin{enumerate}
\item \citet{Bahcall1981} used the two-point correlation method to
  argue that the excess of pairs found at small separations is a
  signature of physically associated pairs; binarity of some of these
  systems was later confirmed by radial velocity observations
  \citep{Latham1984}. See \citet{Garnavich1988} and
  \citet{Wasserman1991} for other studies that use this method. 
\item To reduce the number of false positives inherent in the above,
  one can use additional information such as proper motions. Orbital
  motions for wide systems are small; hence, the space velocities
  of a gravitationally bound pair should be the same, within
  some uncertainty. In the absence of radial velocities, which are
  very hard to obtain for a very large number of field stars, proper
  motion alone can be used to identify binary systems; the resulting
  pairs are known as common proper-motion (CPM) doubles. 
  \citet{Luyten1979a, Luyten1988} pioneered this technique in his
  surveys of Schmidt telescope plates using a blink microscope and
  detected more than 6000 wide CPM doubles with $\mu>$100\masyr\ over
  almost fifty years. This method has since been used to find CPM  
  doubles in the AGK~3 stars by \citet{Halbwachs1986}, in the revised
  New Luyten Two-Tenths \citep[rNLTT;][]{Salim2003} catalog by
  \citet{Chaname2004}, and among the {\em Hipparcos}
  stars in the Lepine-Shara Proper Motion-North \citep[LSPM-N;][]{Lepine2005}
  catalog by \citet{Lepine2007a}. All of these studies use
  magnitude-limited high proper-motion catalogs and, thus, select
  mostly nearby stars.
\end{enumerate}

More recently, \citet*[][hereafter \citetalias{Sesar2008}]{Sesar2008}
searched the SDSS Data Release Six \citep[DR6;][]{Adelman-McCarthy2008} for
CPM binaries with angular separations up to 
30$\arcsec$ using a novel statistical technique that minimizes the
difference between the distance moduli obtained from photometric
parallax relations  for candidate pairs. They matched proper motion
components to within 5\masyr\ and identified $\sim$22000 total candidates 
with excellent completeness, but with a one-third of them expected to
be false positives. They searched the SDSS DR6 catalog for pairs at all
mass ranges and find pairs separated by 2000--47000 AU, at distances
up to 4 kpc.  Similarly, \citet{Longhitano2010} used the angular
two-point correlation function to do a purely statistical study of
wide binaries in the $\sim$675 square degrees centered at the North Galactic
Pole using the DR6 stellar catalog and predicted that there are more
than 800 binaries with physical separations larger than 0.1~pc but
smaller than 0.8~pc. As evidenced by the large false positive rate in
\citetalias{Sesar2008}, such large-scale searches for wide binaries 
generally involve a trade-off between completeness on the one hand and
fidelity on the other, as they depend on statistical arguments for
identification. 

Complementing this type of ensemble approach, a high-fidelity approach may
suffer from incompleteness and/or biases; however, there are a number of
advantages to a ``pure" sample of {\em bona fide} wide binaries such as
that presented in this work. For example, \citet{Faherty2010} searched for CPM
companions around the brown dwarfs in the BDKP catalog \citep{Faherty2009}
and found nine nearby pairs; all of their pairs were followed up
spectroscopically and, hence, have a much higher probability of being
real. As mass, age, and metallicity can all cause variations in the
observed physical properties, e.g.\ in radius or in magnetic activity,
their effects can be very hard to disentangle in a study of single
stars. Components of multiple systems are expected to have been formed
of the same material at the same time, within a few hundred thousand
years of each other (e.g.~\citealt{White2001}; \citealt*{Goodwin2004a};
\citealt{Stassun2008}). Hence, binaries are perfect tools for separating the
effects of mass, age, and metallicity from each other as well 
as for constraining theoretical models of stellar evolution. Some
examples include benchmarking stellar evolutionary tracks
(e.g.~\citealt{White1999}; \citealt*{Stassun2007}; \citealt{Stassun2008}),
investigating the age-activity relations of M dwarfs \citep*[e.g.][]{Silvestri2005},
defining the dwarf-subdwarf boundary for spectral classification
\citep[e.g.][]{Lepine2007b}, and calibrating the metallicity indices
\citep{Woolf2005, Bonfils2005}. Moreover, equal-mass multiples can be
selected to provide identical twins with the same initial
conditions (same mass, age, and metallicity) to explore the intrinsic
variations of stellar properties. In addition, wide binaries (a $>$
100~AU) are expected to evolve independently of each other; even their
disks are unaffected by the distant companion
\citep{Clarke1992}. Components of such systems are effectively two
single stars that share their formation and evolutionary history. In 
essence they can be looked at as {\em coeval laboratories} that can be
used to effectively test and calibrate relations measured for field stars. 
Finally, as interest has grown in detecting exoplanets and in characterizing
the variety of stellar environments in which they form and evolve, 
a large sample of {\em bona fide} wide binaries could provide a rich
exoplanet hunting ground for future missions such as SIM.

In this paper, we present a new catalog of CPM doubles from SDSS, each
with at least one low-mass component, identified by matching proper
motions and photometric distances. In \S~\ref{Sec: observation} we
describe the origin of the input sample of low-mass stars;
\S~\ref{Sec: method} details the binary selection algorithm and the 
construction of a Galactic model built to assess the fidelity of each
binary in our sample. The resulting catalog and its characteristics are 
discussed in \S~\ref{Sec: catalog}. We compare the result of our CPM double 
search with previous studies in \S~\ref{Sec: discussion} and summarize our
conclusions in \S~\ref{Sec: conclusions}.

\section{SDSS Data}\label{Sec: observation}

\subsection{SDSS Sample of Low-Mass Stars}\label{Sec: sample}
SDSS is a comprehensive imaging and spectroscopic survey of the northern
sky using five broad optical bands ($ugriz$) between $\sim$3000--10000\AA\
using a dedicated 2.5 m telescope \citep{York2000, Fukugita1996, Gunn1998}.
Data Release Seven \citep[DR7;][]{Abazajian2009} contains photometry of 357
million unique objects over 11663 $\deg^2$ of the sky and spectra of
over 1.6 million objects over 9380 $\deg^2$ of the sky. The photometry has
calibration errors of 2\% in $u$ and $\sim$1\% in $griz$, with
completeness limits of 95\% down to magnitudes 22.0, 22.2, 22.2, 21.3,
20.5 and saturation at magnitudes 12.0, 14.1, 14.1, 13.8, 12.3,
respectively \citep{Gunn1998}. We restricted our sample to $r\leq20$
where the SDSS/USNO-B proper motions are more reliable (see
\S~\ref{Sec: quality}); hence, photometric quality and completeness 
should be excellent for our sample.

Stellar sources with angular separations $\gtrsim$ 0.5--0.7$\arcsec$ are
resolved in SDSS photometry; we determined this empirically from a
search in the {\sc Neighbors} table. For sources brighter than $r\sim$
20, the astrometric accuracy is 45 mas rms per coordinate while the
relative astrometry between filters is accurate to 25--35 mas rms
\citep{Pier2003}.

\begin{figure*}[htb]
  \begin{centering}
  \includegraphics[width=0.7\linewidth]{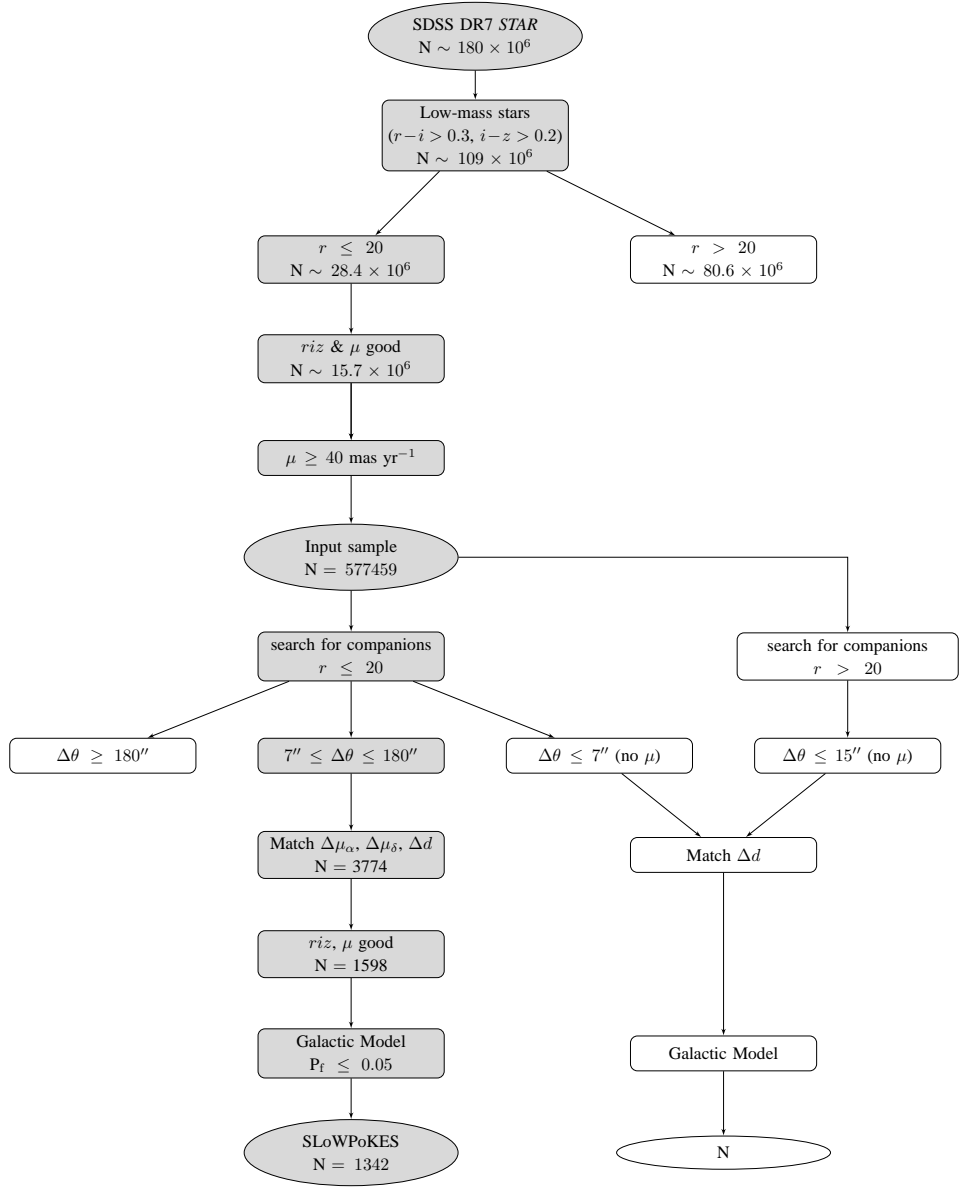}
  \caption{A graphical representation of the selection process used in the
    identification of {\slowpokes} pairs. The gray boxes show the steps
    involved in {\slowpokes}, with the results  presented in this paper. The
    pairs without proper motions (shown in white boxes) will be
    presented in a follow-up paper.}
  \label{Fig: flowchart}
  \end{centering}
\end{figure*}

The SDSS DR7 photometric database has more than 180 million stellar
sources \citep{Abazajian2009}; to select a sample of low-mass stars,
we followed the procedures outlined in \citetalias{Bochanski2010} and
required $r-i \geq$ 0.3 and $i-z \geq$ 0.2, which represents the
locus of sample K5 or later dwarfs. The {\sc star} table was used to
ensure that all of the selected objects had morphology consistent with
being point sources ({\sc type} $=$ 6) and were not duplicate
detections of the same source ({\sc primary})\footnote{We note that
  binaries separated by less than $\sim$1$\arcsec$ might appear
  elongated in SDSS photometry and might not be listed in the {\sc star}
  table.}. This yielded a sample of $>$ 109 million low-mass stars
with $r\sim$ 14--24, at distances of $\sim$0.01--5 kpc from the Sun.
Figure~\ref{Fig: flowchart} shows a graphical flow chart of
the selection process and criteria used; the steps needed in
identifying kinematic companions, which are detailed in this paper,
are shown in gray boxes. The white boxes show subsets where kinematic
information is not available; however, it is still possible to
identify binary companions based on their close proximity. We will
discuss the selection of binaries, without available kinematic
information, in a future paper.  

\subsection{Proper Motions} \label{Sec: pm}
The SDSS/USNO-B matched catalog \citep{Munn2004}, which is integrated
into the SDSS database in the {\sc ProperMotions} table, was used to
obtain proper motions for this study. We used the proper motions from
the DR7 catalog; the earlier data releases had a systematic error
in the calculation of proper motion in right ascension \citep[see][for
details]{Munn2008}. This catalog uses SDSS galaxies to recalibrate the
USNO-B positions and USNO-B stellar astrometry as an additional epoch
for improved proper motion measurements. The {\sc ProperMotions}
catalog is resolution limited in the USNO-B observations to
7$\arcsec$ and is 90\% complete to $g=$ 19.7\footnote{This
  corresponds to $r=$ 18.75 for K5 and $r=$ 18.13 for M6 dwarfs.},
corresponding to the faintness limit of the POSS-II plates used in
USNO-B. The completeness also drops with increasing proper motion; for
the range of proper motions in our sample ($\mu =$ 40--350\masyr;
see \S~\ref{Sec: quality} below), the completeness is $\sim$ 85\%.  The
typical 1 $\sigma$ error is 2.5--5\masyr\ for each component  

\subsection{Quality Cuts}\label{Sec: quality}
To ensure that the resultant sample of binaries is not contaminated
due to bad or suspect data, we made a series of cuts on the stellar
photometry and proper motions. With $>$ 109 million low-mass stars, we
could afford to be very conservative in our quality cuts and still
have a reasonable number of stars in our input sample. We restricted
our sample, as shown in Figure~\ref{Fig: flowchart}, to stars
brighter than $r=$ 20 and made a cut on the standard quality flags on
the $riz$ magnitudes ({\sc peakcenter, notchecked, psf\_flux\_interp,
  interp\_center, bad\_counts\_error, saturated} -- all of which are
required to be 0)\footnote{A {\sc primary} object is already selected
  to  not be {\sc bright} and {\sc nodeblend} or {\sc
    deblend\_nopeak}}, which are the only bands pertinent to low-mass 
stars and the only bands used in our analysis.

On the proper motions, \citet{Munn2004} recommended a minimum total proper
motion of 20\masyr\ and cuts based on different flags for a ``clean''
and reliable sample of stellar sources. Therefore, we
required that each star (i) matched an unique USNO-B source within
1$\arcsec$ ({\sc match} $>$ 0), (ii) had no other SDSS source brighter
than $g=$ 22 within 7$\arcsec$ ({\sc dist22} $>$ 7), (iii) was detected on
at least 4 of the 5 USNO-B plates and in SDSS ({\sc nFit} = 6 or ({\sc
  nFit} = 5 and ({\sc O} $<$ 2 or {\sc J} $<$ 2))), (iv) had a good
least-squares fit to its proper motion ({\sc sigRA} $<$ 1000 and {\sc
  sigDec} $<$ 1000), and (v) had 1 $\sigma$ error for both components
less than 10\masyr.

\begin{figure}
  \begin{centering}
  \includegraphics[width=1\linewidth]{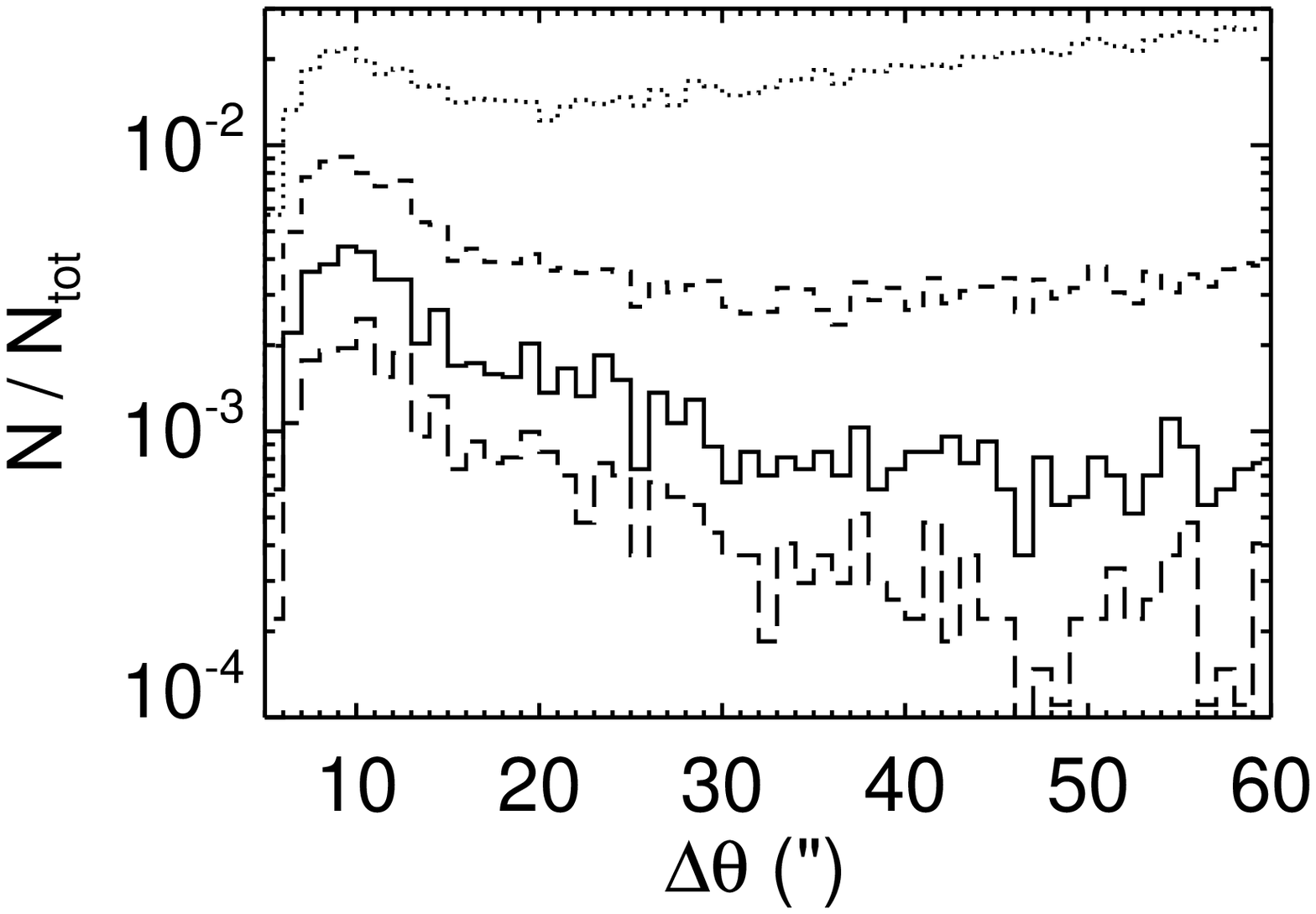}
  \caption{The angular separation distribution of candidate pairs
    with minimum proper motions cut-offs of 20 (dotted), 30 (dashed),
    40 (solid), and 50 (long dashes)\masyr. These candidates were
    selected by matching angular separations, proper motions, and
    photometric distances as described in \S~\ref{Sec: selection}. All
    histograms have been normalized by the area of the histogram with the
    largest area to allow for relative comparisons. It can be clearly
    deduced that a low proper motion cut-off will cause a large
    proportion of chance alignments. We adopted a cut-off of  $\mu
    \geq$ 40\masyr\ as it allows for identification of CPM pairs with
    a reasonable number of false positives that are later sifted with
    our Galactic model.}
  \label{Fig: min_pm}
  \end{centering}
\end{figure}

A challenge inherent in using a deep survey like SDSS to identify CPM
binaries, is that most of the stars are far away and, therefore, have
small proper motions. To avoid confusing real binaries with chance
alignments of stars at large distances, where proper motions are
similar but small, a minimum proper motion cut needs to be
applied. Figure~\ref{Fig: min_pm} shows the distribution of candidate
binaries (selected as detailed in \S~\ref{Sec: selection}) with
minimum proper motion cuts of 20, 30, 40, and 50\masyr; the
histograms have been normalized by the area of the histogram with the
largest area to allow for relative comparisons. All four distributions
have a peak at small separations of mostly real binaries, but the
proportion of chance alignments at wider separations becomes larger
and more dominant at smaller $\mu$ cutoffs. 

If our aim were to identify a complete a sample of binaries, we might
have chosen a low $\mu$ cutoff and accepted a relatively high
contamination of false pairs. Such was the approach in the recent study
of SDSS binaries by \citetalias{Sesar2008} who used $\mu\geq$ 15
\masyr. Our aim, in this paper, was to produce a ``pure'' sample with a high
yield of {\em bona fide} binaries. Thus, in our search for CPM pairs
we adopted a minimum proper motion of $\mu=$ 40\masyr\ for our
low-mass stellar sample since the number of matched pairs clearly declines
with increasing $\dtheta$ with this cutoff (Figure~\ref{Fig: min_pm}). 
While this still allows for a number of (most likely) chance alignments,
they do not dominate the sample and can be sifted more effectively as
discussed below.

\begin{figure*}
  \begin{centering}
  \includegraphics[width=0.8\linewidth]{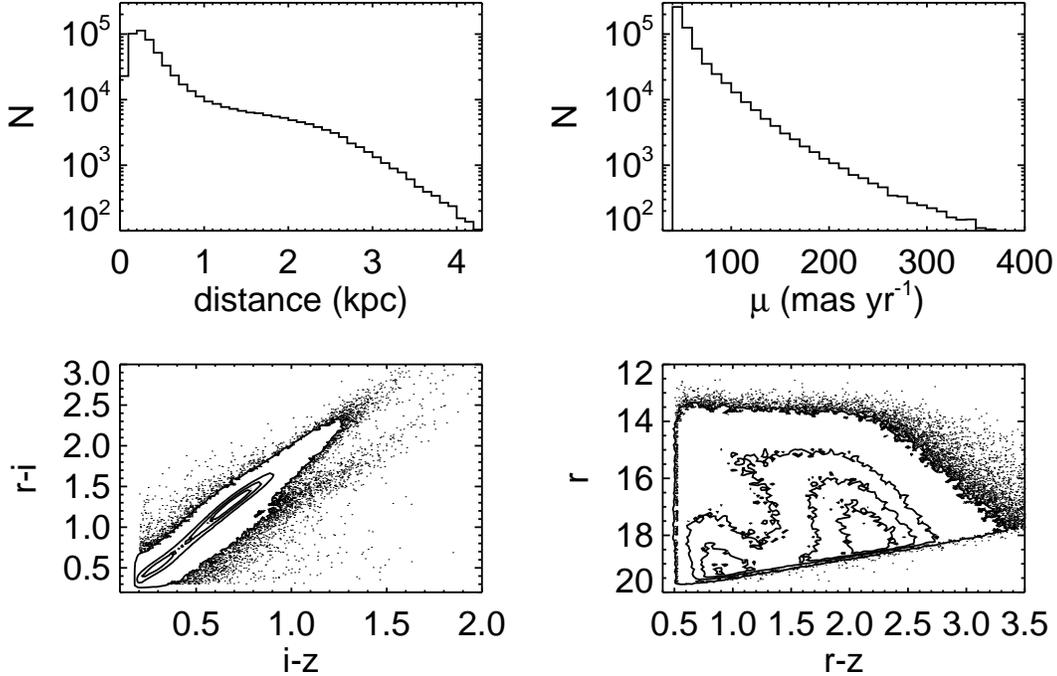}
  \caption{The characteristics of the sample of 577459 low-mass stars
    in the SDSS DR7 photometric catalog that forms our input
    sample {\em (clockwise from top left)}: the 
    photometric distances, the total proper motions, $r$ vs. \rz\ Hess
    diagram, and the $r-i$ vs. $i-z$ color--color plot. In addition to
    quality cuts on proper motion and photometry, we require that
    stars in this sample to be low-mass ($r-i\geq$ 0.3 and $i-z\geq$ 0.2) and
    have a relatively high proper motion ($\mu\geq$ 40\masyr). The
    contours show densities of 10--3000 in the color--color plot and
    10--300 in the Hess diagram.}
  \label{Fig: sample_prop}
  \end{centering}
\end{figure*}

Thus, on the $\sim$15.7 million low-mass stars that satisfied the
quality cuts, we further imposed a $\mu\geq$ 40\masyr\ cut on the
total proper motion that limits the input sample to 577459 stars with
excellent photometry and proper motions. As Figure~\ref{Fig:
  flowchart} shows, this input sample constitutes all the stars
around which we searched for companions. Figure~\ref{Fig: sample_prop}
shows the distributions of photometric distances, total  
proper motions, $r-i$ vs.\ $i-z$ color-color diagram, and the $r$
vs.\ \rz\ Hess diagram for the input sample. As
seen from their \rz\ colors in the Hess diagram, the sample consists
of K5--L0 dwarfs. Metal-poor halo subdwarfs are also clearly
segregated from the disk dwarf population; however, due to a
combination of the magnitude and color limits that were used, they are
also mostly K subdwarfs and are limited to only the earliest spectral
types we probe.

\subsection{Derived properties}
\subsubsection{Photometric Distances} \label{Sec: distance}
\noindent{\bf Disk dwarfs (DD)}:
We determined the distances to the DDs in our sample by using
photometric parallax relations, measured empirically with the SDSS
stars. For M and L dwarfs ($\sim$M0--L0; 0.94 $<$ \rz\ $<$ 4.34),
we used the relation derived by \citetalias{Bochanski2010} based on
data from D.~Golimowski et al. (2010, in prep.).
For higher mass dwarfs ($\sim$O5--K9; $-$0.72 $<$ \rz\ $<$ 0.94),
we fit a third-order polynomial to the data reported in
\citet{Covey2007}:
\begin{equation}%\label{Eqn: phot_plx2}
M_r=3.14+7.54\ (\rz)-3.23\ (\rz)^2+0.58\ (\rz)^3.
\end{equation}
In both of the above relations, we used extinction-corrected
magnitudes. Ideally, $M_r$ would have been a function of both color
and metallicity; but the effect of metallicity is not quantitatively known
for low-mass stars. This effect, along with unresolved binaries and
the intrinsic width of the MS, cause a non-Gaussian scatter of $\sim$
0.3--0.4 magnitudes in the photometric parallax relations
(\citealt*{West2005}; \citetalias{Sesar2008, Bochanski2010}). Since we
are matching the photometric distances,  
using the smaller error bars ensures fewer false matches. Hence, we
adopted 0.3 magnitudes as our error, implying a 1~$\sigma$ error of
$\sim$14\% in the calculated photometric distances.
\\
\\
\noindent{\bf Subdwarfs (SD)}:
Reliable photometric distance relations are not available for SDs, so
instead we used the relations for DDs above. As a result of appearing
under-luminous at a given color, their absolute distances will be
overestimated. However, the {\em relative} distance between two stars in
a physical binary should have a small uncertainty. Because we are
interested in determining if two candidate stars occupy the same
volume in space, the relative distances will suffice. While
  photometric parallax relations based on a few stars of a range of
  metallicities \citep{Reid1998,Reid2001c} or calibrated for
  solar-type stars \citep{Ivezic2008b} do exist and can provide
  approximate distances for low-mass SDs, we refrained from using them due to
  the large uncertainties involved.
\\
\\
\noindent{\bf White dwarfs (WD)}:
We calculated the photometric distances to WDs using the algorithm
used by \citet{Harris2006}: $ugriz$ magnitudes and the $u-g$, $g-r$,
$r-i$, and $i-z$ colors, corrected for extinction, were fitted to the
WD cooling models of \citet*{Bergeron1995} to get the bolometric
luminosities and, hence, the distances\footnote{SDSS $ugriz$
  magnitudes and  colors for the WD cooling models are available on
  P. Bergeron's website: \url{http://www.astro.umontreal.ca/~bergeron/CoolingModels/}}.
The bolometric luminosity of WDs are a function of gravity as well
as the composition of its atmosphere (hydrogen/helium), neither of
which can be determined from the photometry. Therefore, we assumed pure
hydrogen atmospheres with $\log\ g=$ 8.0 to calculate the distances to
the WDs. As a result, distances derived for WDs with unusually low
mass and gravity ($\sim$10\% of all WDs) or with unusually high mass
and gravity ($\sim$15\% of all WDs) will have larger
uncertainties. Helium WDs redder than $g-i>$ 0.3 will also have
discrepant distances \citep{Harris2006}. 

\subsubsection{Spectral Type \& Mass}\label{Sec: SpType}
The spectral type of all (O5--M9; $-0.6 \leq \rz \leq$4.5) disk dwarfs
and subdwarfs were inferred from their \rz\ colors using the following
two-part fourth-order polynomial:
\begin{eqnarray}
  {\rm SpT} = 
  \left\{ \begin{array}{l}
      50.4 + 9.04\ (\rz) - 2.97\ (\rz)^2 + \\
      \quad 0.516\ (\rz)^3 - 0.028\ (\rz)^4 \mbox{; for O5--K5}\\
      33.6 + 47.8\ (\rz) - 20.1\ (\rz)^2 - \\
      \quad 33.7\ (\rz)^3 - 58.2\  (\rz)^4 \mbox{; for K5--M9}
      \end{array}\right.
\end{eqnarray}
where SpT ranges from 0--67 (O0$=$0 and M9$=$67 with all
spectral types, except K for which K6, K8, and K9 are not defined,
having 10 subtypes) and is based on the data reported in
\citet{Covey2007} and \citet{West2008}. The spectral types should be
correct to $\pm$1 subtype. 

Similarly, the mass of B8--M9 disk dwarfs and subdwarfs were
determined from their \rz\ colors using a two-part fourth-order polynomial,:
\begin{eqnarray}
    M(\Msun) =
    \left\{ \begin{array}{l}
        1.21 -1.12\ (\rz) +1.06\ (\rz)^2 - \\
        \quad 2.73\ (\rz)^3 + 2.97\ (\rz)^4 \mbox{; for B8--K5}\\
        0.640 +0.362\ (\rz) -0.539\ (\rz)^2 + \\
        \quad 0.170\ (\rz)^3 -0.016\ (\rz)^4 \mbox{; for K5--M9}
      \end{array}\right.
\end{eqnarray}
based on the data reported in \citet{Kraus2007}, who used theoretical
models, supplemented with observational constrains when needed, to
get mass as a function of spectral type. The scatter of the fit,
as defined by the median absolute deviation, is $\sim$2\%. 

As the polynomials for both the spectral type and mass are monotonic
functions of \rz\ over the entire range, the component with the
bluer \rz\ color was classified as the primary star of each binary found.

\section{Method}\label{Sec: method}
\subsection{Binary Selection} \label{Sec: selection}
Components of a gravitationally bound system are expected to occupy
the same spatial volume, described by their semi-axes, and to move
with a common space velocity. 
To identify physical binaries in our low-mass sample, we implemented a
statistical matching of positional astrometry (right ascension, \ra,
and declination, \dec), proper motion components (\pmra\ and \pmdec),
and photometric distances ($d$). The matching of distances is an
improvement to the methods of previous searches for CPM doubles
\citep{Lepine2007a, Chaname2004, Halbwachs1986} and serves to provide
further confidence in the binarity of identified systems.

The angular separation, \dtheta, between two nearby point sources, A and B,
on the sky can be calculated using the small angle approximation:
\begin{equation}\label{Eqn: ang_sep}
\dtheta \simeq \sqrt{(\alpha_A-\alpha_B)^2 \cos\delta_A \cos\delta_B +
  (\delta_A-\delta_B)^2}.
\end{equation}
We searched around each star in the input sample for all stellar neighbors,
brighter than $r=$ 20 and with good photometry and proper motions, within 
$\dtheta\leq$ 180$\arcsec$ in the SDSS photometric database using the
Catalog Archive Server query
tool\footnote{\url{http://casjobs.sdss.org/CasJobs/}}. Although CPM
binaries have been found at much larger angular separations
(up to 900$\arcsec$ in \citealt{Chaname2004}; 1500$\arcsec$ in
\citealt{Lepine2007a}; 570$\arcsec$ in \citealt{Faherty2010}), the
contamination rate of chance alignments at such large angular
separations was unacceptably high in the deep SDSS imaging (see Figure
\ref{Fig: min_pm}). In addition, searching the large number of matches
at larger separations required large computational resources. However,
since the SDSS low-mass star sample spans considerably larger
distances than in previous studies (see below), our cutoff of 180$\arcsec$
angular separation probes similar {\em physical separations} of up to
$\sim$0.5~pc, which is comparable to the typical size of prestellar
cores \citep*[0.35 pc;][]{Benson1989, Clemens1991, Jessop2000}.  

For all pairs that were found with angular separations of $7\arcsec
\leq \dtheta\leq 180\arcsec$, we required the photometric distances to
be within
\begin{equation}\label{Eqn: ddist}
\Delta d \leq (1~\sigma_{\Delta d}, 100\ {\rm pc})
\end{equation}
and the proper motions to be within
\begin{equation}\label{Eqn: dpm}
\left(\frac{\Delta\pmra}{\sigma_{\Delta\pmra}}\right)^2 +
\left(\frac{\Delta\pmdec}{\sigma_{\Delta\pmdec}}\right)^2 \leq 2
\end{equation}
where $\Delta d$, $\Delta\pmra$, and $\Delta\pmdec$ are the scalar
differences between the two components with
their uncertainties calculated by adding the individual uncertainties
in quadrature. An absolute upper limit on $\Delta d$ of 100~pc was
imposed to avoid $\Delta d$ being arbitrarily large at the
very large distances probed by SDSS; hence, at $d\gtrsim$ 720 pc,
distances are matched to be within 100~pc and results in a
significantly lower number of candidate pairs. The proper
motions are matched in 2-dimensional vector space, instead of just
matching the total (scalar) proper motion as has been frequently done in the
past. The latter approach allows for a significant number of false
positives as stars with proper motions with the same magnitude but
different directions can be misidentified as CPM pairs.

For our sample, the uncertainties in proper motions are almost always
larger than the largest possible Keplerian orbital motions of the
identified pairs. For example, a binary with a separation of 5000~AU
and \Mtot $=$ 1~\Msun\ at 200 pc, which is a typical pair in the
resultant {\slowpokes} sample, will has a maximum orbital motion of
$\sim$0.27~\masyr, much smaller than error in our proper motions
(typically 2.5--5~\masyr). However, a pair with a separation of 500~AU
pair and \Mtot $=$ 1~\Msun\ at 50~pc has a maximum orbital motion of
$\sim$5.62~\masyr, comparable to the largest errors in component
proper motions. Hence, our algorithm will reject such nearby,
relatively tight binaries.

From the mass estimates and the angular separations from the resulting
sample, we calculated the maximum Keplerian orbital velocities, which
are typically less than 1~\masyr; only 104, 34, and 3 out of a total
of 1342 systems exceeded 1, 2, and 5~\masyr, respectively. More
importantly, only 7 pairs had maximum orbital velocities greater than
1-$\sigma$ error in the proper motion; so apart from the nearest
and/or tightest pairs, our search should not have been affected by our
restrictions on the proper motion matching in Eq.~(\ref{Eqn: dpm}).

\begin{figure}
  \begin{centering}
  \includegraphics[width=1\linewidth]{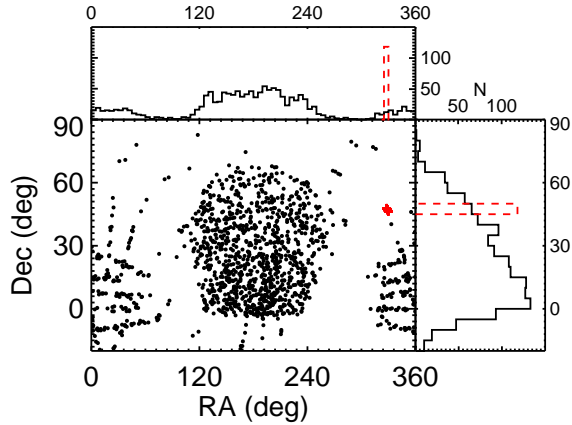}
    \caption{The spatial distribution of {\slowpokes} binaries, shown
      in equatorial coordinates, closely follows the SDSS footprint. 
      The upper and right panels show the histogram of
      Right Ascension and Declination in 5$\degr$ bins. We rejected the
      118 CPM candidates found in the direction of open cluster IC
      5146 (shown in red pluses and dashed histograms) due to the highly
      anomalous concentration of pairs and possibly contaminated
      photometry due to nebulosity (see \S~\ref{Sec: selection}).}
    \label{Fig: radec}
  \end{centering}
\end{figure}

Applying these selection criteria to the stellar sample described in
\S~\ref{Sec: sample}, we found a total of 3774 wide CPM binary
candidates, where each has at least one low-mass component. Among
these, we found 118 pairs shown in red pluses and dashed histograms in
Figure~\ref{Fig: radec}, concentrated in a $20\arcmin\times150\arcmin$
stripe, in the direction of open cluster IC 5146.As
there is significant nebulosity that is not reflected in the
extinction values, the photometry and, hence, the calculated
photometric distances are not reliable. In addition, while the
kinematics are not characteristic of IC 5146, they are more likely to
be part of a moving group rather than individual CPM  pairs. Thus, we
rejected all of these candidates. No other distinct
structures, in space and kinematics, were found. Then, we
made the quality cuts described in \S~\ref{Sec: quality} on the
companions; 906 and 1085 companions did not meet our threshold for the
photometry and proper motions, respectively, and were rejected. The
majority of these rejections were near the Galactic Plane, which was expected
due to higher stellar density. Thus, at the end,
the resulting sample had 1598 CPM double candidates from the
statistical matching (Figure \ref{Fig: flowchart}). Inherent in statistical
samples are false positives, arising from chance alignments within the
uncertainties of the selection criteria. For any star, the probability of
chance alignment grows with the separation making the wider companions
much more likely to be chance alignments despite the selection
criteria we just implemented. Hence, it is necessary to complete a
detailed analysis of the fidelity of the sample.

\subsection{Galactic Model: Assessing False Positives in the Binary 
Sample} \label{Sec: model}

To quantitatively assess the fidelity of each binary in our sample, we
created a Monte Carlo based Galactic model that mimics the spatial and
kinematic stellar distributions of the Milky Way and calculates the
likelihood that a given binary could arise by chance from a random
alignment of stars. Clearly, for this to work the underlying
distribution needs to be carefully constructed such that, as an
ensemble, it reflects the statistical properties of the true
distribution. The model needs to take into account the changes in the
Galactic stellar density and space velocity over the large range of
galactocentric distances and heights above/below the Galactic disk
plane probed by our sample. Previous studies, focused on nearby
binaries, have been able to treat the underlying Galactic stellar
distribution as a simple random distribution in two-dimensional space.
For example, \citet{Lepine2007a} assigned
a random shift of 1--5$\degr$ in Right Ascension for the secondary
of their candidate pairs and compared the resultant distribution
with the real one. As they noted, the shift cannot be arbitrarily
large and needs to be within regions of the sky with similar
densities and proper motion systematics. With stars at much larger
heliocentric distances in our sample, even a 1$\degr$ shift
would correspond to a large shift in Galactic position (1$\degr$ in
the sky at 1000~pc represents $\sim$17.5~pc). Therefore, we could not use
a similar approach to assess the false positives in our catalog.

The Galactic model is built on the canonical view that the
Galaxy comprises three distinct components---the thin disk, the
thick disk, and the halo---that can be cleanly segregated by their
age, metallicity, and kinematics \citep*{Bahcall1980, Gilmore1989,
  Majewski1993}. We note that some recent work has argued for the disk
to be a continuum instead of two distinct components
\citep{Ivezic2008b} or for the halo to be composed of two distinct
components \citep{Carollo2008, Carollo2009}. However, for our purposes, the
canonical three-component model was sufficient. We also did not try 
to model the over-densities or under-densities, in positional or
kinematic space, caused by co-moving groups, open clusters,
star-forming regions, or Galactic streams. If such substructures were
found in the SDSS data, they were removed from our sample (see
\S~\ref{Sec: selection}). In essence, this model strictly describes
stars in the field and produces the three-dimensional position and
two-dimensional proper motion, analogous to what is available for the
SDSS photometric catalog. 

\subsubsection{Galactic stellar density profile} \label{Sec: density_profile}
In the canonical Galactic model, the stellar densities ($\rho$) of the
thin and the thick disks, in standard Galactic coordinates $R$ (Galactic
radius) and $Z$ (Galactic height), are given by
\begin{eqnarray}
\label{Eqn: rho_thin}
  \rho_{\rm thin} (R,Z) &=& \rho\ (R_{\odot},0)\ 
  e^{-\frac{|Z|}{H_{\rm thin}}} \
  e^{-\frac{|R-R_{\odot}|}{L_{\rm thin}}} \\ 
\label{Eqn: rho_thick}
  \rho_{\rm thick}(R,Z) &=& \rho\ (R_{\odot},0)\  
  e^{-\frac{|Z|}{H_{\rm thick}}} \ 
  e^{-\frac{|R-R_{\odot}|}{L_{\rm thick}}},
\end{eqnarray}
where $H$ and $L$ represent the scale height above (and below) the
plane and the scale length within the plane, respectively. The
halo is described by a bi-axial power-law ellipsoid
\begin{equation}
\label{Eqn: rho_halo}
  \rho_{\rm halo} (R,Z) = \rho\ (R_{\odot},0) 
  \left(\frac{R_{\odot}}{\sqrt{R^2+(Z/q)^2}}\right)^{r_{\rm halo}}
\end{equation}
with a halo flattening parameter $q$ and a halo density gradient
$r_{\rm halo}$. The three profiles are added together, with the
appropriate scaling factors, $f$, to give the stellar density profile
of the Galaxy: 
\begin{equation}\label{Eqn: rho_total}
  \rho\ (R,Z) = 
  f_{\rm thin}\  \rho_{\rm thin} + 
  f_{\rm thick}\ \rho_{\rm thick} +
  f_{\rm halo}\  \rho_{\rm halo}.
\end{equation}
The scaling factors are normalized such that
$f_{\rm thin} + f_{\rm thick} + f_{\rm halo} = 1$. With the large
number of stars imaged in the SDSS, robust stellar density functions
have been measured for the thin and thick disks using the low-mass
stars (\citealt{Juric2008}; \citetalias{Bochanski2010}) and for the halo using the
main-sequence turn-off stars \citep{Juric2008}. The values measured
for the disk in the two studies are in rough agreement. We adopted the
disk parameters from \citetalias{Bochanski2010} and the halo
parameters from \citet{Juric2008}; Table~\ref{Tab: gal_model}
summarizes the adopted values. 

% ************** Table: GALACTIC MODEL ***************
\begin{center}
\begin{deluxetable}{lllc}
\tablewidth{0pt}
\tablecaption{Galactic Structure Parameters}
\tablehead{
  \colhead{Component}    & \colhead{Parameter name} & 
  \colhead{Parameter description} & \colhead{Adopted Value}}
\startdata
           & $\rho\ (R_{\odot}, 0)$ & stellar density & 0.0064 \\
\hline
           & $f_{\rm thin}$  & fraction\tablenotemark{a} & 1-$f_{\rm thick}$-$f_{\rm halo}$ \\
thin disk  & $H_{\rm thin}$  & scale height     & 260~pc \\ 
           & $L_{\rm thin}$  & scale length     & 2500~pc \\
\hline
           & $f_{\rm thick}$ & fraction\tablenotemark{a} & 9\% \\
thick disk & $H_{\rm thick}$ & scale height     & 900~pc \\
           & $L_{\rm thick}$ & scale length     & 3500~pc \\
\hline
           & $f_{\rm halo}$  & fraction\tablenotemark{a} & 0.25\% \\
halo       & $r_{\rm halo}$  & density gradient & 2.77 \\
           & $q\ (=c/a)\tablenotemark{b}$    & flattening parameter & 0.64 \\            
\enddata
\label{Tab: gal_model}
\tablenotetext{a}{Evaluated in the solar neighborhood}
\tablenotetext{b}{Assuming a bi-axial ellipsoid with axes {\em a} and {\em c}}
\tablecomments{The parameters were measured using M dwarfs for the
  disk \citep{Bochanski2010} and main-sequence turn-off stars for the
  halo \citep{Juric2008} in the SDSS footprint.}
\end{deluxetable}
\end{center}

\subsubsection{Galactic Kinematics} \label{Sec: kinematics}
Compared to the positions, the kinematics of the stellar components
are not as well characterized; in fact, apart from their large
velocity dispersions, little is known about the halo
kinematics. We seek to compare the proper motions of a candidate
pair with the expected proper motions for that pair given its Galactic
position. Thus, we found it prudent to (i) ignore the halo component,
with its unconstrained kinematics, at distances where its
contributions are expected to be minimal and (ii) limit our model to a
distance of 2500~pc, which corresponds to the Galactic height where
the number of halo stars begins to outnumber disk stars
\citep{Juric2008}. In practice, all the {\slowpokes} CPM pairs, with
the exceptions of subdwarfs for which distances were known to be
overestimated, were within $\sim$1200~pc (see Figure \ref{Fig:
  bin_grid} below); so we did not introduce any significant
systematics with these restrictions.

An ensemble of stars in the Galactic Plane can be characterized as
having a purely circular motion with a velocity, $V_{\rm circ}$. The
orbits become more elliptic and  
eccentric over time due to kinematic heating causing the azimuthal
velocity, $V_{\phi}$, to decrease with the Galactic height,
Z. However, the mean radial ($V_r$) and perpendicular ($V_z$) velocities
for the ensemble at any $Z$ remains zero, with a given dispersion, as
there is no net flow of stars in either direction. In addition, this
randomization of orbits also causes the asymmetric drift, $V_{\rm a}$, which
increases with the age of stellar population and is equivalent to
$\sim$10 \kms\ for M dwarfs. Hence, the velocities of stars in the
Galactic disk can be summarized, in Galactic cylindrical coordinates, as:
\begin{eqnarray}\label{Eqn: vel_mu}
\langle V_r~(Z) \rangle &=& 0 \nonumber \\
\langle V_{\phi}~(Z) \rangle &=& V_{\rm circ} - V_{\rm a} - f(Z)\\
\langle V_z~(Z) \rangle &=& 0\nonumber,
\end{eqnarray}
where $V_{\rm circ}=$ 220 \kms, $V_{\rm a}=$ 10 \kms, $f(Z)=0.013~|Z| - 1.56
\times10^{-5}~|Z|^2$ was derived by fitting a polynomial to the 
data in \citet{West2008}, and $Z$ is in parsecs. This formulation of 
$V_{\phi} (Z)$ is consistent with a stellar population composed of a faster
thin disk ($\langle V_{\phi}\rangle =$ 210 \kms) and a slower thick disk
($\langle V_{\phi}\rangle =$ 180 \kms). Then, we converted these
galactocentric polar velocities to the heliocentric, Cartesian {\em UVW}
velocities. The {\em UVW} velocities, when complemented with the 
dispersions, can be converted to a two-dimensional proper motion
\citep[and radial velocity;][]{Johnson1987}, analogous to our input 
catalog. We used the {\em UVW} velocity dispersions measured for SDSS
low-mass dwarfs \citep{Bochanski2007a}. All the dispersions were well
described by the power-law 
\begin{equation}\label{Eqn: vel_disp}
\sigma(Z) = k\ |Z|^n,
\end{equation}
where the values of constants $k$ and $n$ are summarized in Table
\ref{Tab: UVW}. As the velocity dispersions in \citet{Bochanski2007a}
extend only up to $\sim$1200 pc, we extrapolated the above equation
for larger distances. While the velocity ellipsoids of F
and G dwarfs have been measured to larger distances
\citep[e.g.][]{Bond2009}, we preferred to use the values measured for M
dwarfs for our low-mass sample.

% ************** Table: GALACTIC KINEMATICS ***************
\begin{center}
\begin{deluxetable}{lcrc}
\tablewidth{0pt}
\tablecaption{Galactic Kinematics}
\tablehead{
  \colhead{Galactic component} & \colhead{Velocity} & 
  \colhead{{\em k}} & \colhead{{\em n}}}
\startdata
           & {\em U} & 7.09 & 0.28 \\
thin disk  & {\em V} & 3.20 & 0.35 \\
           & {\em W} & 3.70 & 0.31 \\
\hline
           & {\em U} & 10.38 & 0.29 \\
thick disk & {\em V} & 1.11  & 0.63 \\
           & {\em W} & 0.31  & 0.31 \\
\enddata
\label{Tab: UVW}
\tablecomments{The constants in the power law, $\sigma (Z) = k |Z|^n$,
  that describes the velocity dispersions of the stars in the thin and
  thick disks. The velocity dispersions were measured from a
  spectroscopic sample of low-mass dwarfs \citep{Bochanski2007a}.}
\end{deluxetable}
\end{center}

\subsubsection{The Model}
By definition, a chance alignment occurs because two physically
unassociated stars randomly happen to be close together in our line of
sight (LOS), within the errors of our measurements. Due to the random
nature of these chance alignments, it is {\em not} sufficient to
estimate the probability of chance alignment along a given LOS simply
by integrating Eq.~(\ref{Eqn: rho_total}). This would tend to
underestimate the true number of chance alignments because the density
profiles in Eqs.~(\ref{Eqn: rho_thin}--\ref{Eqn: rho_halo}) are smooth
functions that, in themselves, do not include the random scatter about
the mean relation that real stars in the real Galaxy have. Thus, the
stars need to be randomly redistributed spatially about the average
given by Eq.~(\ref{Eqn: rho_total}) in order to properly account for
small, random fluctuations in position and velocity that could give
rise to false binaries in our data.

In principle, one could simulate the whole Galaxy in this fashion in
order to determine the probabilities of chance alignments as a
function of LOS. In practice, this requires exorbitant 
amounts of computational time and memory. Since our aim was to calculate
the likelihood of a false positive along specific LOSs, we, instead,
generated stars in much smaller regions of space, corresponding to the
specific binaries in our sample. For example, a
$30\arcmin\times30\arcmin$ cone integrated out to a distance of 2500
pc from the Sun will contain at least a few thousand stars within any
of the specific LOSs in our sample. The number of stars is large
enough to allow for density variations similar to that of the Milky
Way while small enough to be simulated with ease. With a sufficient
number of Monte Carlo realizations, the random density fluctuations
along each LOS can be simulated. We found that $10^5$ realizations
allowed for the results to converge, within $\sim$0.5\%.

We implemented the following recipe to assess the false-positive likelihood
for each candidate pair using our Galactic
model:
\begin{enumerate}
\item The total number of stars in the LOS volume defined by
  a $30\arcmin\times30\arcmin$ area, centered on the \ra\ and \dec\ of a
  given binary, over heliocentric distances of 0--2500~pc was
  calculated by integrating Eq.~(\ref{Eqn: rho_total}) in 5~pc deep,
  discrete cylindrical ``cells.'' 

  Integrating Eq.~(\ref{Eqn: rho_total}) for \dtheta\ $=$ 30$\arcmin$
  and $d=$ 0--2500~pc resulted in $\sim$3300--1580 stars, with the higher
  numbers more likely to be the LOSs along the Galactic Plane. This number of
  stars, when randomly redistributed in the entire volume, was more
  than enough to recreate over-densities and under-densities.

\item The stars were then distributed in three-dimensional space defined by
  the LOS using the rejection method \citep{Press1992}, generating
  \ra, \dec, and $d$ for each star. The rejection method ensured that
  the stars were randomly distributed while following the overlying
  distribution function, which, in this case, was the stellar density
  profile given by Eq.~(\ref{Eqn: rho_total}).

  The model did an excellent job of replicating the actual
  distribution of stars in the three-dimensional space. The red
  histograms in Figure~\ref{Fig: model_density} show the number of stars
  from the center of the LOS as a function of angular separation for
  the model (dashed lines) and the data (solid lines), averaged over
  all LOSs where candidate binaries were identified in \S~\ref{Sec:
    selection}. There is an excess of pairs at close separations, a
  signature of genuine, physically associated pairs, while the two
  distributions follow the same functional form at larger
  separations, where chance alignments dominate. The increasing number
  of pairs with angular separation is an evidence of larger volume
  that is being searched. Integrating the model predicts $\sim$8.0
  stars within a search radius of 60$\arcsec$ and $\sim$71 within
  180$\arcsec$. In other words, for a typical LOS with candidate binaries,
  we would, on average, expect to find 8 chance alignments within
  60$\arcsec$ and $\sim$71 within 180$\arcsec$ when considering only
  angular separation.

  For the 180$\arcsec$ radius around each LOS, the average number of
  stars in the model and the data were within a few $\sigma$ of each
  other, with the largest deviations found along the LOSs at low
  Galactic latitudes or large distances. As our model only integrated
  to $d=$ 2500~pc, deviations for LOSs at large distances were
  expected. Deviations for LOSs at low Galactic latitudes are
  reasonable as the parameters in Eqs.~(\ref{Eqn: rho_thin}) and
  (\ref{Eqn: rho_thick}) as not as well constrained along the Galactic disk.
  Hence, we concluded that the rejection method used in redistributing
  the stars in the $30\arcmin\times30\arcmin$ LOS is correctly
  implemented and that the model successfully replicates the three
  dimensional spatial distribution of the stars in the Galaxy.
  
  When the distances were matched, the number of pairs decreased as
  chance alignments were rejected (blue histograms in Figure~\ref{Fig:
    model_density}); {\em there are, on average, only $\sim$0.41 and
    $\sim$3.6 chance alignments within 60$\arcsec$ and 180$\arcsec$,
    respectively.} Note that matching distances, in addition to the
  angular separation, significantly enhanced the peak at the small
  separations.
  
  \begin{figure}
    \begin{centering}
      \includegraphics[width=1\linewidth]{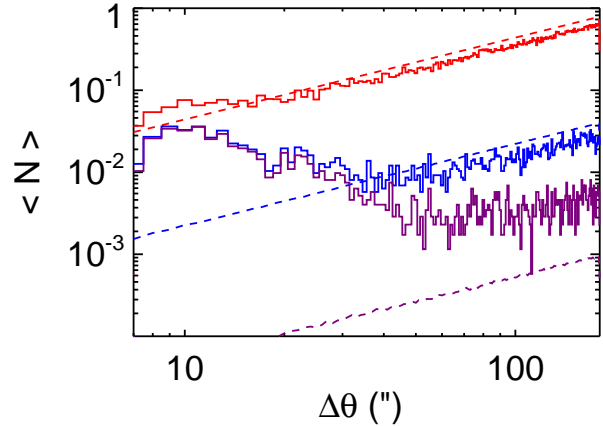}
      \caption{The distribution of the (average) number of stars found
        around the LOSs around our binary candidates as calculated from the SDSS
        DR7 data (solid histograms) and our Galactic model (dashed
        histograms). All optical pairs (red), pairs with matching
        distances (blue), and pairs with matching distances and proper
        motion components (purple) are shown. The excess at small
        separations, a signature of real pairs, is enhanced as additional
        properties are matched. Unlike the rest of the paper, we
        counted all stars with $r\leq$ 22.5, without any quality cuts,
        for this plot to do a realistic comparison with the model. Note
        that $\langle N \rangle$ denotes the number of stars found at
        that angular separation (counted in 1$\arcsec$ bins). We used the
        raw parameters for the Galactic model
        (\citealt{Juric2008}, \citetalias{Bochanski2010}).
        As a whole, the model does an excellent job of
        mimicking  the number of stars and their spatial and kinematic
        distribution in the Milky Way along typical SDSS LOSs.}
      \label{Fig: model_density}
    \end{centering}
  \end{figure}

\item Based on the Galactic position of each randomly generated star, 
  mean {\em UVW} space velocities and their dispersions were generated based on
  Eq.~(\ref{Eqn: vel_mu}) and Eq.~(\ref{Eqn: vel_disp}),
  respectively. Proper motions were then calculated by using the
  inverse of the algorithm outlined in \citet{Johnson1987}. These
  generated proper motions represent the expected kinematics of stars
  at the given Galactic position.

  Figure~\ref{Fig: model_pm} shows the comparison between the proper
  motions in the SDSS/USNO-B catalog and our model; component proper
  motions of stars within 60$\arcmin$ of all LOSs, where candidate
  pairs were found. For the purposes of this plot, we restricted the
  stars to be within 1200~pc and of spectral type K5 or later so we
  could compare kinematics with a sample similar to our resultant
  catalog. We also did not compare the distributions at \pmra\ or
  \pmdec\ $<$ 10\masyr where the proper motions are comparable to the
  1-$\sigma$ errors and, hence, not reliable. Since our initial sample
  was has $\mu\geq$ 40\masyr, stars with the small
  proper motion component are either rejected or their motion is
  dominated by the other component in our analysis. As evidenced in
  the figure, our model reproduced the overall kinematic structure of
  the thin and thick disks.

  \begin{figure}
    \begin{centering}
      \includegraphics[width=1\linewidth]{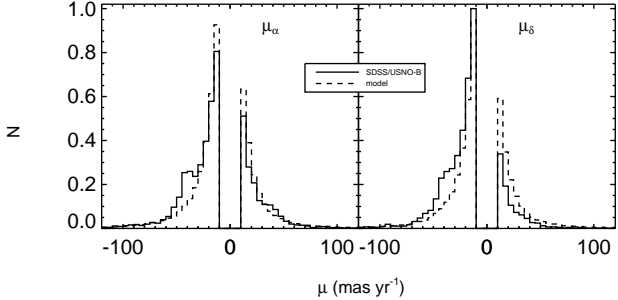}
      \caption{Proper motion distributions for stars within 60$\arcmin$ of
        LOSs along all identified candidate pairs in SDSS/USNO-B (solid
        histograms) and in our
        Galactic model (dashed histograms). For the purposes of this plot,
        we restricted the stars to be of spectral types K5 or later and be
        within 1200~pc to avoid being skewed by systematic differences. In
        addition, we do not compare proper motion components $<$ 10\masyr\
        as they comparable to the 1-$\sigma$ errors and, hence, 
        not reliable (see text). Again, the kinematics of the thin and
        thick disk of the Milky Way are very well reproduced by our Monte
        Carlo model.}
      \label{Fig: model_pm}
    \end{centering}
  \end{figure}

  When the proper motions were matched (purple histograms in Figure
  \ref{Fig: model_density}), in addition to the angular separation and
  distances, the number of chance alignments fell drastically,
  especially at the smaller separations. In fact, at \dtheta\ $\leq$
  15$\arcsec$, there were $< 10^{-4}$ chance alignments, on average;
  even at \dtheta $=$ 180$\arcsec$, the real pairs outnumbered the chance
  alignments by a factor of 4--5. Cumulatively, for a typical LOS,
  there were, on average, only $\sim$0.0097 and $\sim$0.086 chance
  alignments within 60$\arcsec$ and 180$\arcsec$, respectively. As a 
  result of matching distance and proper motions components, the
  number of chance alignments within 180$\arcsec$ were reduced by 
  a factor of $\sim$800.
  
\item In the model galaxy, we repeated the selection process used
  to find CPM pairs in the SDSS photometric catalog, as described in
  \S~\ref{Sec: selection}. To avoid double-counting, the input
  coordinates of the LOS were considered to be the primary star. Note
  that we did not intend to model and reproduce both stars of a given
  pair but wanted to see if a random chance alignment could produce a
  companion for a given primary. Hence, each additional star that was
  found to satisfy Eqs.~(\ref{Eqn: ddist}) and (\ref{Eqn: dpm}) was
  counted as an optical 
  companion. The average number of companions found in the $10^5$
  Monte Carlo realizations is the probability of chance alignment or
  the probability that the candidate pair is a false positive,
  \Pf, for that candidate pair. The number of realizations sets the
  resolution of \Pf\ at 10$^{-5}$. 

\item Finally, this was repeated for all candidate pairs that were
  found in \S~\ref{Sec: selection}. Figure~\ref{Fig: P_f} shows the
  distribution of \Pf\ all of the candidate pairs; we further discuss
  this distribution is \S~\ref{Sec: fidelity}.
\end{enumerate}

To conclude, the result of our Galactic model was a 
$30\arcmin\times30\arcmin$ area of the sky, centered around the
given CPM pair, with the surrounding stars following the Galactic
spatial and kinematic distributions of the Milky Way. Each star in
this model galaxy was described by its position (\ra, \dec,
and $d$) and proper motion (\pmra and \pmdec), same as is
available for SDSS photometric catalog. Based on the above results, we
concluded that the Galactic model sufficiently reproduced the
five-dimensional (three spatial and two kinematic\footnote{The model
  also predicts radial velocities. We have an observational program
  underway to obtain radial velocities of the binaries for further
  refinement of the sample.}) distribution of stars along typical LOSs
in the Milky Way and, thus, allowed for the calculation of probability
that a given CPM double is a chance alignment.

\subsection{Fidelity}\label{Sec: fidelity}
As described in the previous section,
we have implemented a very stringent selection algorithm in identifying the
CPM pairs. In addition to only including objects with the most robust
photometry and proper motions, we also used a relatively high
proper motion cut of $\mu \geq$ 40\masyr\ for the input sample,
which considerably decreased the number of low-mass stars. As described 
above, an algorithm optimized to reject the most false positives, even at the
expense of real pairs, was used in the statistical matching of angular
separation, photometric distance, and proper motion
components. Lastly, we used the Galactic model to quantify the
probability of chance alignment, \Pf, for each of the 1558 candidate pairs. 

Normally all candidate pairs with \Pf\ $\leq$ 0.5, i.e.\ a
higher chance of being a real rather than a fake pair, could be used
to identify the binaries. However, to minimize the number of spurious
pairs, we required
\begin{equation}\label{Eqn: Pf}
\Pf \leq 0.05. \\
\end{equation}
for a pair to be classified as real. Here we note that \Pf\ represents
the false-alarm probability that a candidate pair identified by
matching angular separation, photometric distance, and proper motion
components, as described by Eqs.~(\ref{Eqn: ddist}) and (\ref{Eqn:
  dpm}), is a real pair; it is not the probability that a random low-mass
star is part of a wide binary. 

\begin{figure}
  \begin{centering}
  \includegraphics[width=1\linewidth]{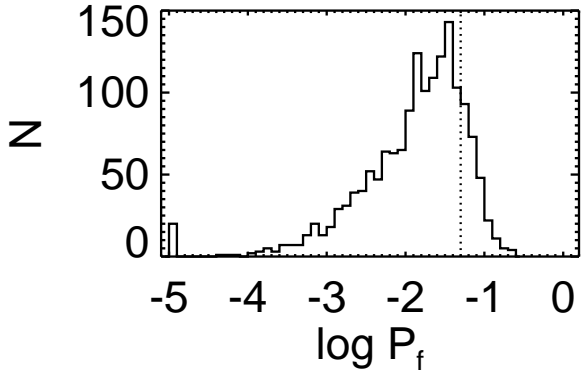}
  \caption{The probability of chance alignment (\Pf)
    for candidate {\slowpokes} pairs based on their positions
    and kinematics, as calculated by our Galactic model. We adopted
    \Pf\ $=$ 0.05 (dotted line) as our threshold for inclusion in the
    {\slowpokes} catalog. The resolution of the Monte Carlo simulation is
    $10^{-5}$, which causes the peak at $\log$ \Pf\ $= 10^{-5}$.}
  \label{Fig: P_f}
  \end{centering}
\end{figure}

Making the above cut on \Pf\  resulted in a catalog of 1342 pairs,
with a maximum of 5\% or 67 of the pairs expected to be false
positives. However, as a large number of the pairs have 
extremely small \Pf\ (see Figure~\ref{Fig: P_f}),
the number of false positives is likely to be much smaller. Adding
up the \Pf\ for the pairs included in the catalog gives an estimated
22 (1.65\%) false positives. In other words, the overall fidelity of
{\slowpokes} is 98.35\%. This is a remarkably small proportion for a sample of very
wide pairs, especially since they span a large range in heliocentric
distances and is a testament to our selection criteria. For example, if the 
proper motion components were matched to within 2~$\sigma$, Eq.~(\ref{Eqn:
  Pf}) would have rejected $\gtrsim$ 60\% of the candidates. Our choice
of Eq.~(\ref{Eqn: Pf}) is a matter of preference; if a more efficient
(or larger) sample is needed, it can be changed to suit the purpose.

To get a first-order approximation of how many real binaries we are
missing or rejecting, we applied our selection algorithm to the rNLTT
CPM catalog with 1147 pairs \citep{Chaname2004}. Note that
\citet{Chaname2004} did not match distances of the components, as they
did not have reliable distance estimates available and matched total
proper motions instead of a two-dimensional vector matching in our
approach. Out of the 307 rNLTT pairs, which have $\dtheta \leq 180\arcsec$
and are within the SDSS footprint, we recover both components of 194 systems
(63\%) and one component of another 56 (18\%) systems, within
2$\arcsec$ of the rNLTT coordinates. Of the 194 pairs for which both 
components had SDSS counterparts, 59 (30\%) pairs satisfied our criteria for
proper motions, Eq.~(\ref{Eqn: dpm}), while 19 (10\%) pairs satisfied
our criteria for both proper motions and distances, Eqs.~(\ref{Eqn:
  ddist}) and (\ref{Eqn: dpm}). In other words with our selection
criteria, we recovered only 10\% of the rNLTT pairs, with SDSS
counterparts, as real binaries. If we relax our selection 
criteria to match within 2-$\sigma$, the number of recovered pairs
increases to 82 with matching proper motions and 37 with matching
proper motions and distances. Of course, as the rNLTT is a nearby,
high proper motion catalog, \citet{Chaname2004} had to allow for
larger differences in proper motions due to the stars' orbital motion,
which is not applicable to the {\slowpokes} sample (see \S~\ref{Sec:
  selection}). In conclusion, compared to previous catalogs of CPM
doubles, we find a small fraction of previously identified very wide
binaries. The reasons for the low recovery rate are two-fold: (i) the 
restrictive nature of our matching algorithm that rejects the most false
positives, even at the expense of real pairs and (ii) improvement in
the identification method---e.g.\ matching proper motions in vector
space and being able to use photometric distance as an additional criterion.

% ************** Table: slowpokes ***********
\begin{center}
\begin{deluxetable*}{rrrrrccccccc}[hb]
\tablecolumns{12}
\tablecaption{Properties of {\slowpokes} pairs}
\tablehead{
  \colhead{ID\tablenotemark{a}} &
  \multicolumn{4}{c}{Position (J2000)}&
  \colhead{ } &
  \multicolumn{6}{c}{Photometry\tablenotemark{b}}\\
  \cline{2-5}
  \cline{7-12}
  \colhead{ } &
  \colhead{$\ra_A$}&\colhead{$\dec_A$} &
  \colhead{$\ra_B$}&\colhead{$\dec_B$} &
  \colhead{ } & 
  \colhead{$r_A$}& \colhead{$i_A$}& \colhead{$z_A$} &
  \colhead{$r_B$}& \colhead{$i_B$}& \colhead{$z_B$} \\
  \colhead{SLW} &  
  \multicolumn{4}{c}{(deg)}&
  \colhead{ } &
  \multicolumn{6}{c}{(mag)}}
\startdata
 J0002+29 &    0.515839 &   29.475183 &    0.514769 &   29.470617 && 16.79 (0.02) & 15.85 (0.02) & 15.33 (0.02) & 19.35 (0.02) & 17.91 (0.02) & 17.17 (0.02) \\ 
 J0004-10 &    1.122441 &  -10.324043 &    1.095927 &  -10.296753 && 18.25 (0.02) & 17.08 (0.02) & 16.46 (0.02) & 18.66 (0.02) & 17.42 (0.02) & 16.70 (0.02) \\ 
 J0004-05 &    1.223125 &   -5.266612 &    1.249632 &   -5.237299 && 14.85 (0.01) & 14.31 (0.01) & 14.02 (0.02) & 19.56 (0.02) & 18.10 (0.01) & 17.31 (0.02) \\ 
 J0005-07 &    1.442631 &   -7.569930 &    1.398478 &   -7.569359 && 17.35 (0.02) & 16.25 (0.01) & 15.65 (0.01) & 18.78 (0.02) & 17.43 (0.02) & 16.69 (0.01) \\ 
 J0005+27 &    1.464802 &   27.325805 &    1.422484 &   27.300282 && 18.91 (0.02) & 17.63 (0.02) & 16.94 (0.01) & 19.79 (0.02) & 18.29 (0.02) & 17.50 (0.02) \\ 
 J0006-03 &    1.640868 &   -3.928988 &    1.641011 &   -3.926589 && 17.06 (0.01) & 16.01 (0.02) & 15.44 (0.01) & 17.88 (0.01) & 16.62 (0.02) & 15.93 (0.01) \\ 
 J0006+08 &    1.670973 &    8.454040 &    1.690136 &    8.498541 && 19.48 (0.02) & 18.14 (0.02) & 17.45 (0.02) & 19.57 (0.02) & 18.15 (0.01) & 17.43 (0.02) \\ 
 J0007-10 &    1.917002 &  -10.340915 &    1.924510 &  -10.338869 && 17.43 (0.01) & 16.50 (0.01) & 16.02 (0.03) & 18.33 (0.01) & 17.24 (0.01) & 16.66 (0.03) \\ 
 J0008-07 &    2.135590 &   -7.992694 &    2.133660 &   -7.995245 && 17.33 (0.01) & 16.35 (0.02) & 15.86 (0.02) & 18.14 (0.01) & 16.98 (0.02) & 16.36 (0.02) \\ 
 J0009+15 &    2.268841 &   15.069630 &    2.272453 &   15.066457 && 18.52 (0.02) & 17.00 (0.02) & 16.09 (0.01) & 18.39 (0.02) & 16.83 (0.02) & 15.95 (0.01) \\ 
\enddata
\end{deluxetable*}
\end{center}

\begin{center}
\begin{deluxetable*}{rrrrcrrcrrcrccrrr}[hb]
\tablecolumns{17}
\tablewidth{0pt}
\tablenum{3}
\tablecaption{}
\tablehead{
  \multicolumn{4}{c}{Proper Motion} &  
  \colhead{ } &
  \multicolumn{2}{c}{Distance\tablenotemark{c}} &
  \colhead{ } &
  \multicolumn{2}{c}{Spectral Type\tablenotemark{d}} &
  \colhead{ } &
  \multicolumn{6}{c}{Binary Information}\\ 
  \cline{1-4}
  \cline{6-7}
  \cline{9-10}
  \cline{12-17}
  \colhead{$\mu_{\ra_A}$}&\colhead{$\mu_{\dec_A}$}&
  \colhead{$\mu_{\ra_B}$}&\colhead{$\mu_{\dec_B}$}&
  \colhead{ } &
  \colhead{$d_A$} & \colhead{$d_B$} &
  \colhead{ } &
  \colhead{$A$} & \colhead{$B$} &
  \colhead{ } &
  \colhead{\dtheta}&\colhead{\dmu}&\colhead{$\Delta d$} &
  \colhead{BE} &\colhead{\Pf} & \colhead{Class\tablenotemark{e}}\\
  \multicolumn{4}{c}{(\masyr)}&
  \colhead{ } &
  \multicolumn{2}{c}{(pc)}&
  \colhead{ } &
  \multicolumn{2}{c}{ }&
  \colhead{ } &
  \colhead{($\arcsec$)} & \colhead{(\masyr)} & \colhead{(pc)} &
  \colhead{($10^{40}$~ergs)} & \colhead{(\%)} & \colhead{}}
\startdata
   198 (2) &     38 (2) &    197 (3) &     35 (3) &&  341 &  301 && M1.7 & M3.6 &&  16.8 &  2.9 &  39 &   58.08 &  0.000 & SD \\ 
    43 (3) &     -4 (3) &     36 (4) &     -8 (4) &&  362 &  338 && M2.7 & M3.1 && 135.9 &  7.5 &  23 &    3.94 &  0.036 & DD \\ 
   101 (2) &     11 (2) &     99 (4) &      8 (4) &&  301 &  333 && K7.1 & M3.8 && 142.0 &  3.1 &  31 &   13.45 &  0.006 & DD \\ 
    30 (2) &    -21 (2) &     30 (3) &    -27 (3) &&  296 &  273 && M2.4 & M3.4 && 157.6 &  6.0 &  23 &    4.90 &  0.015 & DD \\ 
    10 (3) &    -40 (3) &      5 (4) &    -40 (4) &&  357 &  380 && M3.1 & M3.9 && 163.6 &  5.5 &  22 &    2.26 &  0.037 & DD \\ 
   -40 (2) &    -35 (2) &    -40 (2) &    -31 (2) &&  242 &  270 && M2.2 & M3.1 &&   8.7 &  3.9 &  27 &  112.47 &  0.005 & DD \\ 
   -48 (4) &     -5 (4) &    -41 (4) &     -3 (4) &&  417 &  476 && M3.3 & M3.5 && 174.1 &  7.5 &  58 &    1.59 &  0.033 & DD \\ 
    37 (3) &     34 (3) &     42 (4) &     31 (4) &&  429 &  445 && M1.5 & M2.3 &&  27.6 &  5.6 &  16 &   27.74 &  0.034 & DD \\ 
    -9 (2) &    -42 (2) &     -7 (3) &    -41 (3) &&  344 &  379 && M1.7 & M2.7 &&  11.5 &  2.3 &  34 &   74.69 &  0.017 & DD \\ 
    36 (3) &    -16 (3) &     37 (3) &    -21 (3) &&  150 &  161 && M4.2 & M4.2 &&  17.0 &  4.8 &  11 &   21.02 &  0.004 & DD \\ 
\enddata
\tablenotetext{a}{The identifiers were generated using
  the standard {\em Jhhmm$\pm$dd} format using coordinates of the
  primary star and are prefaced with the string 'SLW'.}
\tablenotetext{b}{All magnitudes are psfmag and have not been
  corrected for extinction. Their errors are listed in
  parenthesis. Note that we use extinction-corrected magnitudes in
  our analysis.}
\tablenotetext{c}{The distances were calculated using photometric parallax
  relations and have 1\ $\sigma$ errors of $\sim$14\%. The absolute
  distances to subdwarfs (SDs) are overestimated (see \S \ref{Sec: distance}).}
\tablenotetext{d}{The spectral types were inferred from the \rz\ colors
  \citep{West2008,Covey2007} and are correct to $\pm$1 subtype.}
\tablenotetext{e}{Class denotes the various types of pairs in
  {\slowpokes}. See Table~\ref{Tab: SLW3}.}

\tablecomments{The first 10 pairs are listed here; the full version of
  the table is available online.}
\label{Tab: SLW}
\end{deluxetable*}
\end{center}

\section{Characteristics of the {\slowpokes} catalog} \label{Sec: catalog}

Using statistical matching of angular separation, photometric
distance, and proper motion components, we have identified 1598 very wide,
CPM double candidates from SDSS DR7. We have built a Galactic model, based on 
empirical measurements of the stellar density profile and kinematics, to
quantitatively evaluate the probability that each of those candidate
doubles are real (Figure~\ref{Fig: P_f}). Using Eq.~(\ref{Eqn: Pf}),
we classify 1342 pairs as real, associated pairs. In deference to
their extremely slow movement around each other, we dub the
resulting catalog {\slowpokes} for Sloan Low-mass Wide Pairs of
Kinematically Equivalent Stars. Table~\ref{Tab: SLW} lists the
properties of the identified pairs. The full catalog is publicly
available on the world wide 
web\footnote{\url{http://www.vanderbilt.edu/astro/slowpokes/}}.  
{\slowpokes} is intended to be a ``live'' catalog of wide, low-mass pairs,
i.e., it will be updated as more pairs are identified and as follow-up 
photometric and spectroscopic data become available.

\begin{figure*}
\begin{centering}
  \includegraphics[width=0.8\linewidth]{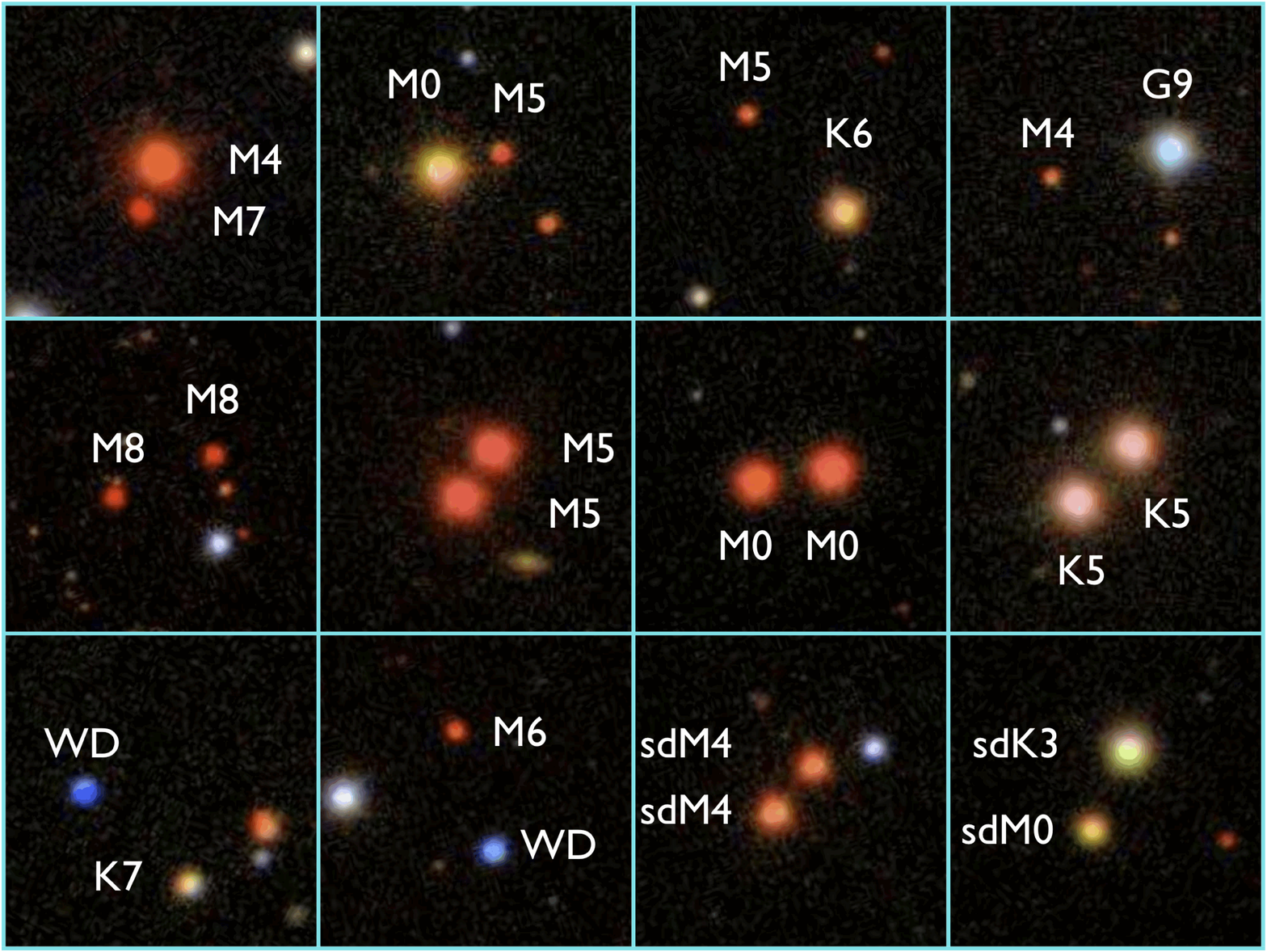}
    \caption{$50\arcsec \times 50\arcsec gri$ composite images of
      CPM pairs found in the {\slowpokes} survey. Pictured are
      high-mass ratio pairs (top row), identical twins (middle row), white
      dwarf--disk dwarf pairs (bottom row, left), and halo subdwarf
      pairs (bottom row, right). Spectral types, based on their \rz\
      colors are shown. Overall, 1342 wide, low-mass binaries were
      identified.}
    \label{Fig: collage}
  \end{centering}
\end{figure*}

Figure~\ref{Fig: collage} shows a collage of $gri$ composite
images, 50$\arcsec$ on a side, of a selection of representative
{\slowpokes} systems; the images are from the SDSS database. In the
collage high-mass ratio pairs (top row; mass ratio = $m_2/m_1 < 0.5$),
equal-mass pairs (middle row; having masses within $\sim$5\% of each
other), white dwarf--disk dwarf pairs (bottom row, left), and halo
subdwarf pairs (bottom row, right) are shown. Table~\ref{Tab: SLW3}
summarizes the different types of systems in the catalog.

% ************** Table: bin_types ***************
\begin{center}
\begin{deluxetable}{clcc}
\tablewidth{0pt}
\tablecaption{The {\slowpokes} binaries}
\tablehead{\colhead{Class} & \colhead{Type} & \colhead{Number}}% & \colhead{Note}}
\startdata
DD  & disk dwarf    & 1245   \\
SD  & subdwarf & 70  \\
WD  & white dwarf--disk dwarf    & 21   \\
T   & triple    & 4     \\
\enddata
\label{Tab: SLW3}
\end{deluxetable}
\end{center}

In the sections that follow, we summarize various aspects of the 
systems that constitute the {\slowpokes} catalog and briefly examine 
some of the follow-up science that can be pursued with {\slowpokes}.
We wish to emphasize that the {\slowpokes} catalog is intended principally
as a high-fidelity sample of CPM doubles that can be used for a 
variety of follow-up investigations where the reliability of each object
in the catalog is more important than sample completeness. Thus, we
have not attempted to fully account for all sources of incompleteness 
or bias; and we intentionally have not applied any form of statistical 
``corrections" to the catalog.

We note that some of the most important sources of incompleteness in the
present catalog could be at least partially overcome with follow-up 
observations. For example, the principal incompleteness in {\slowpokes} arises 
from the lack of proper motions for SDSS stars that were not detected in 
USNO-B. Proper motions either do not exist or are not reliable for stars (i) 
fainter than $r=$ 20 or (ii) within 7$\arcsec$ of a brighter star. The first
criterion currently rules out most of the mid--late M dwarf
companions, while the latter rejects close binaries and most
hierarchical higher-order systems. However, as the SDSS photometric
and astrometric data are available for these systems, 
their multiplicity could be verified with more rigorous analysis or
through cross-matching with other catalogs. For example, by cross-matching 
SDSS with 2MASS, an M4.5--L5 binary \citep{Radigan2008} and an M6--M7 binary
\citep{Radigan2009} have already been identified. At the other end 
of the spectrum, SDSS saturates at $r \approx$ 14, resulting in saturated 
or unreliable photometry for the brighter stars. We found that more than 1000
candidate pairs were rejected for this reason; with reliable follow-up 
photometry these could be added as additional genuine {\slowpokes} binaries.

\begin{figure}
  \begin{centering}
  \includegraphics[width=1\linewidth]{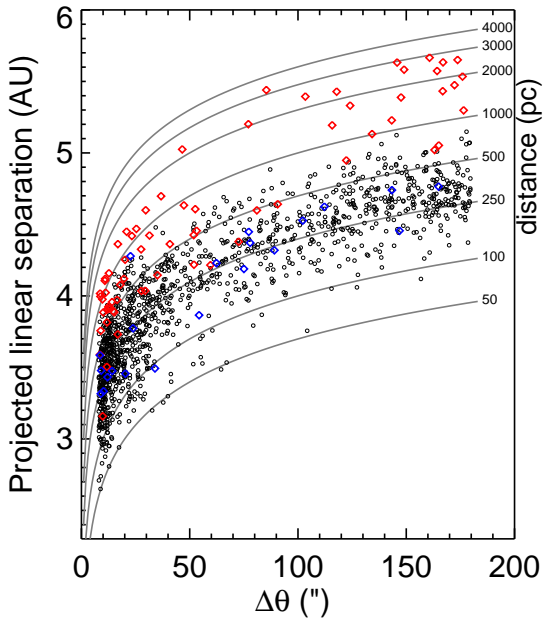}
    \caption{The projected linear separation of {\slowpokes} pairs are
      plotted as a function of angular separation, with the
      iso-distance lines plotted across the grid. DD (black), SD
      (red), and WD--DD pairs (blue) are shown. Systematically larger
      distances for the SDs is clearly seen, which is expected as
      photometric DD photometric parallax relations were used to
      calculate the SD distances as well. Hence, SDs are excluded in
      all analysis involving their distances.}
    \label{Fig: bin_grid}
  \end{centering}
\end{figure}

Even so, a fully volume-complete sample is likely to be impossible to compile
over the full ranges of spectral types and distances spanned by our catalog.
For example, our magnitude limits of 14 $\lesssim r \leq$ 20 imply that
while we are sensitive to K5 dwarfs at $\sim$250--3900~pc, we are sensitive
to M5 dwarfs at $\sim$14--180~pc. This means (i) we are entirely
insensitive to pairs with the most extreme mass ratios (i.e.\ a K5
paired with an M6 or M7) and (ii) we cannot directly compare the
properties of K5 and M5 spectral subtypes in identical distance ranges. 
In addition, as illustrated in Figure~\ref{Fig: bin_grid},
with our 7$\arcsec \leq \dtheta \leq$ 180$\arcsec$ search radius
we are sensitive to companions with separations of $\sim$700--18000~AU at 
100~pc but $\sim$7000--180000~AU at 1000~pc. Thus, it is important with the 
current catalog that statistical determinations of ensemble system properties 
be performed within narrowly defined slices of separation and distance. We
do so in the following subsections as appropriate; but we emphasize again
that our intent here is primarily to {\em characterize} the {\slowpokes}
sample and will proffer any interpretive conclusions only tentatively.

Finally, as is evident from Table~\ref{Tab: SLW3}, the current {\slowpokes}
catalog is clearly not well suited for study of higher-order multiples (i.e.\
triples, quadruples, etc.). Identifying CPM higher-order multiple
systems in SDSS is very challenging due to  
the lack of reliable proper motions in the SDSS/USNO-B matched catalog at
$\dtheta < 7\arcsec$ and $r>$ 20. Unless all components are widely
separated and are all bright, they will be rejected in our
search. In addition, we are currently rejecting hierarchical triples
consisting of a close pair that is unresolved in SDSS and a wide, CPM
tertiary. If the mass ratio of the close, unresolved pair is near
unity, it will appear as an over-luminous single star that will then be
misinterpreted by our algorithm as having a discrepant photometric
distance from its wide tertiary companion.
The available SDSS photometry and astrometry shows evidence of a substantial 
number of such multiple systems, and we plan to make these the subject
of a future study. Already, four CPM triples are identified in our search
(Table~\ref{Tab: SLW3}). Moreover, the current {\slowpokes} catalog is likely
to contain quadruple systems in which the two components of the identified
wide binary are themselves in fact spatially unresolved binaries with near-equal
mass components. We have initiated an adaptive optics program to identify
such higher-order systems in the {\slowpokes} catalog.

\subsection{Kinematic Populations} \label{Sec: pop}
\citet{Luyten1922} devised the reduced proper motion (RPM) diagram to
be used in the stead of the H-R diagram when distances to
the objects are not available, as is the case in large imaging
surveys. The RPM of an object is defined as
\begin{equation}
{\rm H} \equiv m + 5~\log\mu + 5  = M + 5~\log v_{\rm t} - 3.25
\end{equation}
where $v_{\rm t}$ is the heliocentric tangential velocity in \kms\ given by
$v_{\rm t}=4.74~\mu~d$ and $\mu$ is the proper motion in
arcseconds~yr$^{-1}$. Just as in a H-R diagram, the RPM diagram effectively 
segregates the various luminosity or kinematic classes from each other
(e.g.\ \citealt{Chaname2004, Harris2006, Lepine2007a}; \citetalias{Sesar2008}). 

\begin{figure}
  \begin{centering}
  \includegraphics[width=1\linewidth]{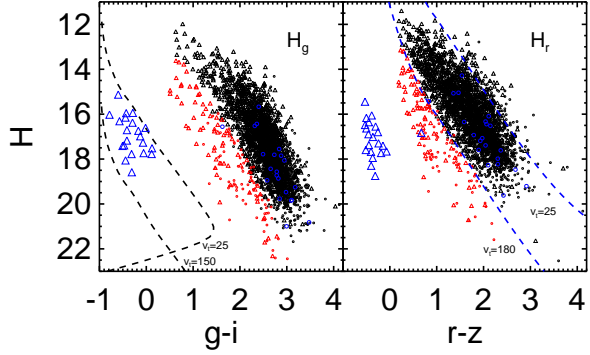}
    \caption{{\slowpokes} pairs plotted in the H$_g$ vs. $g-i$ (left) and
      H$_r$ vs. \rz\ RPM diagrams; primaries (triangles) and secondaries
      (circles) of DD (black), SD (red), and WD--DD (blue) pairs are
      shown. The WDs are identified from the $g$-band RPM diagram,
      complemented by $M_g(g-i)$ for WDs, as having 25 $<v_{\rm t}<$ 150
      \kms\ (black dashed lines) and are all expected to be part of
      the disk population. SDs are segregated from the DDs using the
      $r$-band RPM diagram, assuming SDs have $v_{\rm t}>$ 180
      \kms\ (blue dashed line) and the DD photometric parallax relations.}
    \label{Fig: rpm1}
  \end{centering}
\end{figure}

WDs, in addition to being relatively very blue, are less luminous than
either the DDs or the SDs; hence, the observed WDs tend to be nearby
disk WDs with high tangential velocities. Specifically, spectroscopic
follow-up has shown that the disk WDs with $v_{\rm t}=$ 25--150~\kms\
(Figure~\ref{Fig: rpm1}, black dashed lines) can be effectively 
identified from the $g$-band RPM diagram, when complemented by photometric
parallax relations \citep{Kilic2006}. However, we note that
spectroscopic follow-up is needed to confirm that the identified
objects are actually WDs. The available SDSS spectra confirm that 9
of the 21 WD primaries, identified in SLoWPoKES, are indeed
WDs. As the WDs were identified from the $g$-band, they scatter toward the
SD and DD locus in the $r$-band RPM diagram.

Subdwarfs are low-metallicity halo counterparts of the MS dwarfs
found in the Galactic disk. Hence, they have bluer colors at a given
absolute magnitude \citep[however, the $g-r$ colors for M subdwarfs are
redder;][]{West2004, Lepine2008a} and have higher velocity
dispersions. As a result, the subdwarfs lie below the disk dwarfs in the RPM
diagram\footnote{\citetalias{Sesar2008} showed that the RPM diagram
  becomes degenerate at $Z>$ 2--3 kpc due to the decrease in
  rotational velocity. This does not affect our sample.}. To segregate
the SDs from the DDs, we used the DD photometric parallax
relations complemented by the mean tangential velocity of halo
stars, $v_{\rm t}=$ 180~\kms. This relation is shown in blue
dashed lines in the $r$-band RPM diagram in Figure~\ref{Fig: rpm1}. As
the mean halo velocity was used, SDs can scatter above the line; but
DDs would not be expected to be below the blue line. For comparison,
$v_{\rm t}=$ 25~\kms, the mean tangential velocity of disk stars, is
also shown. Note that DDs in {\slowpokes} have tangential velocities
  larger than the mean velocity of the disk, which is expected as we
  rejected all stars with $\mu<$ 40~\masyr.

\begin{figure}
  \begin{centering}
  \includegraphics[width=1\linewidth]{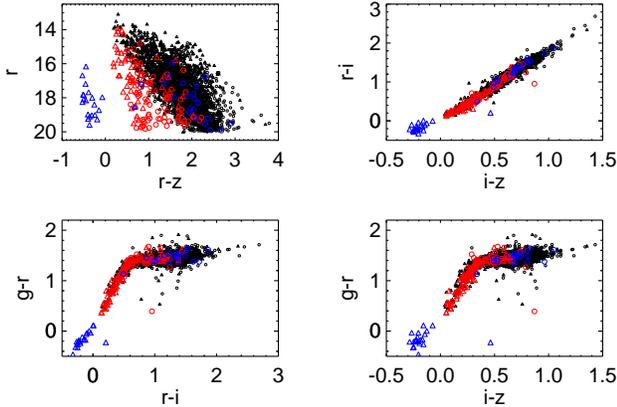}
    \caption{$r$ vs. \rz\ Hess diagram and color-color plots for
      the {\slowpokes} catalog; shown are the primary (triangles) and
      secondaries (circles) for DD (black), SD (red), and WD--DD pairs
      (blue). The subdwarfs are mostly of K spectral type and do not show
      redder $g-r$ colors at given $r-i$ or $i-z$ colors, as seen in M
      subdwarfs \citep{West2004, Lepine2008a}.} 
    \label{Fig: hess}
  \end{centering}
\end{figure}

Figure~\ref{Fig: rpm1} shows the H$_g$ vs.\ $g-i$ and H$_r$ vs.\ \rz\
RPM diagrams with both the primary (triangle) and the secondary
component (circle) for the 21 WD--DD (blue), 70 SD--SD (red), and
1245 DD--DD (black) pairs that were identified in {\slowpokes}. The
properties of the DDs, SDs, and WDs in various color-magnitude and
color-color planes are compared in Figure~\ref{Fig: hess}. The
identified SDs are either K or early-M spectral types, as our
magnitude and color limits exclude the M subdwarf locus. The 
overestimation in the distances to SDs is clearly evident in 
Figure~\ref{Fig: bin_grid}, as they are at systematically larger
distances relative to the DDs. As a result, the calculated physical
separations for the SDs are also systematically larger. At present
a substantial number of subdwarf candidates are rejected from
{\slowpokes} because the overestimated distances result in $\Delta d >$
100~pc.

\begin{figure}
  \begin{centering}
  \includegraphics[width=0.8\linewidth]{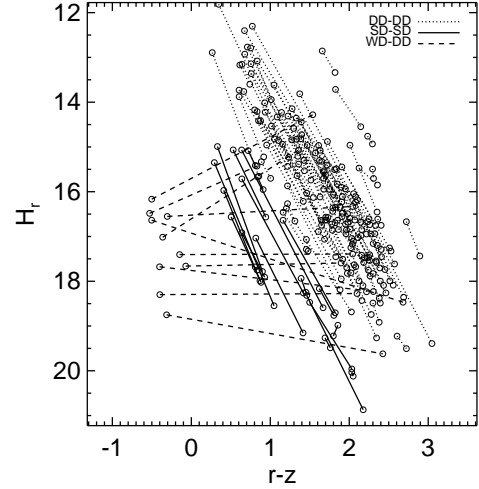}
    \caption{A randomly selected sample of {\slowpokes} pairs
      plotted on the RPM diagram; components of DD (dotted), SD
      (solid), and WD--DD (dashed) pairs are connected. As binaries
      are expected to be part of the same kinematic populations, they
      are expected to lie parallel to the dwarf/subdwarf tracks with
      the obvious exception of WD--DD pairs or when the two components
      lie within the error bars of each other.}
    \label{Fig: rpm2}
  \end{centering}
\end{figure}

RPM diagrams have been used to confirm the binarity of CPM pairs in
the past \citep[e.g.][]{Chaname2004}. As components of a binary system
most probably formed from the same material and at the same time, they
should be members of the same luminosity class (with the obvious
exception of WD--DD pairs) and the line joining the components should
be parallel to the track in which the systems reside. In the case of
WD--DD dwarf systems or systems that have separations comparable to
the error bars in the RPM diagram, the line need not be parallel. The
$r$-band RPM diagram, in Figure~\ref{Fig: rpm2}, confirms that
the {\slowpokes} systems are associated pairs.

\subsection{Separation}
\begin{figure}
  \begin{centering}
  \includegraphics[width=1\linewidth]{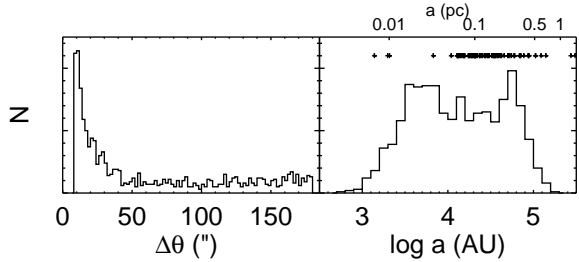}
    \caption{The projected angular separation (left) and the
      semi-major axes \citepalias[right; corrected using $a=1.26\ \theta\ 
      d$,][]{Fischer1992} for the CPM doubles identified in
      {\slowpokes}. The upper x-axis in the right panel shows the
      semi-major axis in parsecs; the widest CPM pairs known in the
      literature (see Table~\ref{Tab: wide_binaries}) are shown as
      pluses at their semi-major axis values at the top of the 
      panel. SDs were excluded from the right panel. The distribution
      of physical separations is clearly bimodal with a break at $\sim$0.1~pc.}
    \label{Fig: bin_sep}
  \end{centering}
\end{figure}

Despite the large number of optical companions found at larger
angular separations, the final distribution of the identified pairs is
mostly of pairs with small angular separations. This is shown in
Figure~\ref{Fig: bin_sep} where, after a narrow peak at
$\dtheta\sim15\arcsec$, the distribution tapers off and rises gently
after $\dtheta\gtrsim 70\arcsec$. To convert the angular separations
into physical separations, we need to account for the projection
effects of the binary orbit. As that information for each CPM pair is not
available, we apply the statistical correction between projected
separation ($s$) and true separation ($a$) determined by
\citetalias{Fischer1992} from Monte Carlo simulations over a full
suite of binary parameters. They found that 
\begin{equation}
a \approx 1.26\ s = 1.26\ \dtheta\ d
\end{equation}
where the calculated $a$ is the physical separation including corrections
for both inclination angle and eccentricity of the binary orbit and is
the actual semi-major axis. We emphasize, however, that these $a$
values are valid only for ensemble comparisons and should not be
taken as an accurate measure of $a$ for any individual system. In
addition, the above equation implies that, at the extrema of
angular separations probed, we are biased towards systems that are
favorably oriented, either because of $\sin i$ projection effects or
eccentricity effects that lead to a changing physical separation as a
function of orbital position.  For example, at \dtheta\ $\approxeq
7\arcsec$, pairs at their maximal apparent separation are more likely
to be identified while we are biased towards pairs with smaller
apparent separations at \dtheta\ $\approxeq 180\arcsec$.

The distribution of $a$ is plotted for the {\slowpokes} sample 
in Figure~\ref{Fig: bin_sep} (right). The semi-major axes for the
identified pairs range from $\sim10^3-10^5$  AU ($\sim$0.005--0.5~pc),
with sharp cutoffs at both  
ends of the distribution, and with a clear bimodal structure in between.
The cutoff at small separations is observational, resulting from our bias
against angular separations $\dtheta < 7\arcsec$. Indeed, the range of 
physical separations probed by {\slowpokes} is at the tail-end of the 
log-normal distribution ($\langle a \rangle=$ 30~AU, $\sigma_{\log
  a}=$ 1.5) proposed by \citetalias{Duquennoy1991} and
\citetalias{Fischer1992}. The cutoff we observe at the other end of
the distribution, $a \sim 10^5$~AU, is likely physical.
With the mean radii of prestellar cores observed to be around 0.35~pc 
or $10^{4.9}$~AU \citep{Benson1989, Clemens1991, Jessop2000}, {\slowpokes} 
represents some of the widest binaries that can be formed. However,
other studies have found binary systems at similar separations;
plotted in pluses are the wide CPM pairs compiled from the
literature and listed in Table~\ref{Tab: wide_binaries}.

% ************** Table: wide_binaries ***************
\begin{center}
\begin{deluxetable*}{llrrllrc}
\tablewidth{0pt}
\tabletypesize{\scriptsize}
\tablecaption{A sub-sample of previously known wide binaries with the
  projected separation, $s$, $\gtrsim 10^4$~AU ($\sim$0.05~pc)}
\tablehead{
  \multicolumn{2}{c}{ID\tablenotemark{a}} &
  \colhead{\dtheta} & \colhead{$s$} &
  \multicolumn{2}{c}{Spectral Type\tablenotemark{b}} &
  \colhead{BE\tablenotemark{c}} & \colhead{References}\\
  \cline{1-2}
  \cline{5-6}
  \colhead{Primary} & \colhead{Secondary} & \colhead{(\arcsec)} & \colhead{(AU)} &
  \colhead{Primary} & \colhead{Secondary}  & \colhead{($10^{40}$~ergs)} & \colhead{ }}
\startdata
   HIP   38602    &     LSPM  J0753+5845 &  109.90 &   10175 &    G8.1 &    M5.1 &    132.18 &        1 \\
   HIP   25278    &      HIP       25220 &  707.10 &   10249 &    F1.8 &    K5.8 &    307.44 &        1 \\
   HIP   52469    &      HIP       52498 &  288.00 &   10285 &    A8.9 &    F9.3 &    379.74 &        1 \\
   HIP   86036    &      HIP       86087 &  737.90 &   10400 &    F8.4 &    M2.6 &    184.33 &        1 \\
   HIP   58939    &     LSPM  J1205+1931 &  117.20 &   10558 &    K0.5 &    M4.7 &    107.44 &        1 \\
   HIP   51669    &     LSPM  J1033+4722 &  164.30 &   10668 &    K7.3 &    M5.7 &     52.10 &        1 \\
   HIP   81023    &     LSPM J1633+0311N &  252.00 &   10723 &    K3.2 &    M3.8 &     81.51 &        1 \\
   HIP   50802    &     LSPM  J1022+1204 &  311.90 &   11139 &    K5.9 &    M3.1 &     56.44 &        1 \\
   HIP   78128    &     LSPM  J1557+0745 &  144.60 &   11385 &    G7.1 &    M5.0 &    123.01 &        1 \\
   HIP  116106    &    2MASS  J2331-04AB &  451.00 &   11900 &      F8 & \nodata &    161.10 &        2 \\
$\alpha$ Cen AB   &     NLTT       37460 & 9302.00 &   12000 &   G2+K1 &      M6 &    430.21 &        3 \\
 \enddata
\tablenotetext{a}{We have tried to use HIP and NLTT identifiers,
  whenever they exist, for consistency.}
\tablenotetext{b}{Spectral types are from the referenced papers,
  SIMBAD, or inferred from their $V-J$ colors using
  \citet{Kenyon1995}.}
\tablenotetext{c}{Binding energies are calculated using estimated
  masses as a function of spectral type \citep{Kraus2007}. When
  spectral type for the secondary was not available, it was assumed to
  be an equal-mass binary.}
\tablerefs{
(1) \citet{Lepine2007a};
(2) \citet{Caballero2007b}; 
(3) \citet{Caballero2009}; 
(4) \citet{Caballero2010}; 
(5) \citet{Bahcall1981};
(6) \citet{Latham1984};
(7) \citet{Makarov2008}; 
(8) \citet{Osorio2004}; 
(9) \citet{Faherty2010}; 
(10) \citet{Chaname2004}; 
(11) \citet{Quinn2009}; 
(12) \citet{Poveda2009};
(13) \citet{Allen2000}.}
\tablecomments{The first 10 pairs are listed here; the full version of
  the table is available online.}
\label{Tab: wide_binaries} 
\end{deluxetable*}
\end{center}

Between the cutoffs at $\sim 10^3$~AU and at $\sim 10^5$~AU, we observe
a distinct bimodality in the distribution of physical separations 
at $a \sim 10^{4.2}$~AU ($\sim$0.1 pc) that has no correlation with
the distance to the observed system (see \S~\ref{Sec: discussion} below).
We have high confidence that this bimodality is not due to some sort
of bias in our sample. As discussed above, the most important observational 
bias affecting the distribution is the bias against pairs 
with $a \lesssim 10^3$~AU, because we are not sensitive to systems with 
$\dtheta < 7\arcsec$ nor to very nearby systems. In addition,
given the care with which we have eliminated false positives in the sample,
we have high confidence that the bimodal structure is 
not due to a large contamination of very wide chance pairs.
Instead, it is likely that this bimodality reveals two distinct 
populations of wide binaries in {\slowpokes}, possibly representing 
systems that form and/or dissipate through differing mechanisms.
We discuss the bimodality in the context of models of binary formation
and kinematic evolution in \S~\ref{Sec: discussion}. 

\subsection{Mass Distribution} \label{Sec: massdist}
\begin{figure}
  \begin{centering}
  \includegraphics[width=1\linewidth]{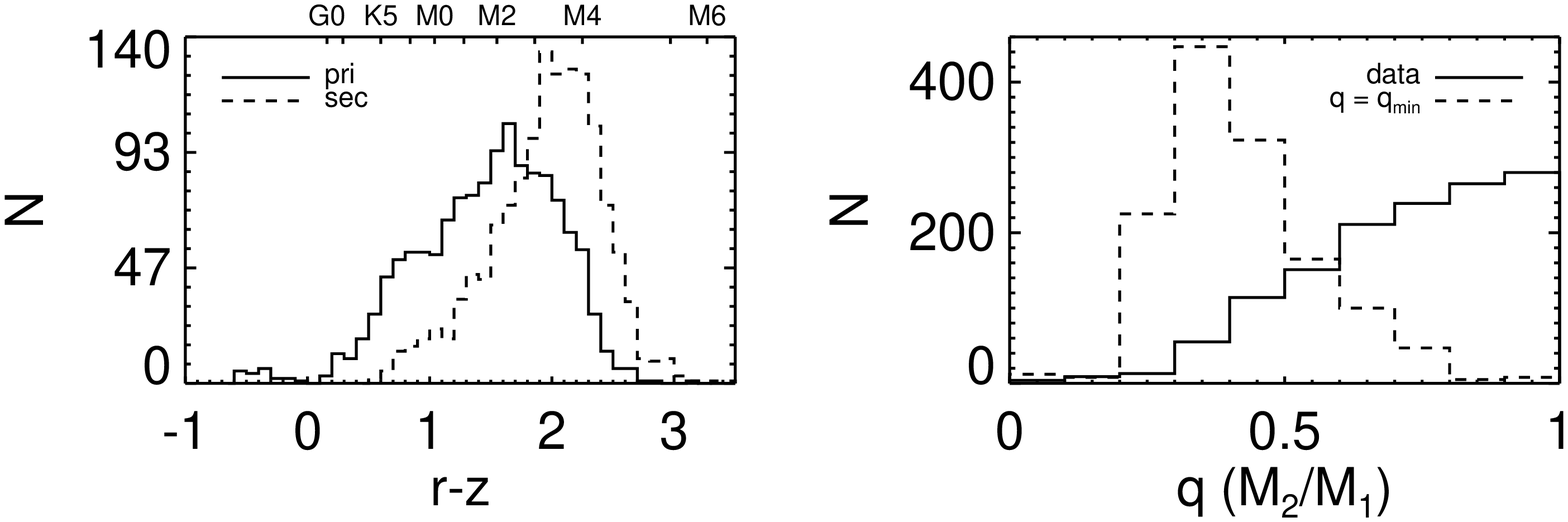}
    \caption{{\em Left}: The \rz\ color for the primaries (solid) and
      secondaries (dashed). The upper x-axis, in the left panel, shows
      the spectral types inferred from their \rz\ colors \citep{Covey2007,
        West2008}. Both distributions exhibit bimodality and are,
      similar to each other.
      {\em Right}: The mass ratio, $q\equiv M_2/M_1$ for {\slowpokes}
      pairs (solid histogram) indicates that most are in equal- or
      near-equal-mass systems. Within the observational limits of
      SDSS, we should have been able to see much more extreme mass
      ratio systems (dashed histogram). The lack of such systems
      indicates that the observed distribution is not dominated by
      observational biases and is probably real.}
    \label{Fig: mass1}
  \end{centering}
\end{figure}

Figure~\ref{Fig: mass1} (left panel) shows the \rz\ color
distributions of the primary and secondary components of {\slowpokes}
pairs, where the primary is defined as the component with the bluer
\rz\ color (and, thus, presumably more massive). As the {\slowpokes} sample
is dominated by dwarfs, the \rz\ color distribution should correspond
directly to the mass distribution. Both the primary and secondary
distributions show a peak at the early--mid M spectral types (as
inferred from their colors; \S~\ref{Sec: SpType}), probably due to, at
least in part, the input sample being mostly M0--M4 dwarfs. However,
apart from the saturation at $r \lesssim$ 14, there is no bias against
finding higher-mass companions. Hence, it is notable that more than
half of the primaries have inferred spectral types of M0 or later,
with the distribution of the secondaries even more strongly skewed to
later spectral types (by definition).

\begin{figure}
  \begin{centering}
  \includegraphics[width=1\linewidth]{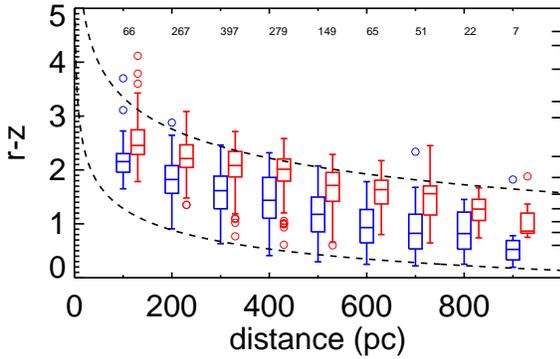}
    \caption{The color distribution for the primary (blue) and
      secondary (red) components of {\slowpokes} pairs as a function of
      distance, in 100~pc bins shown in a box-and-whisker plot; the
      number of pairs in each bin is printed along the top. The
      bar inside the boxes show the median of the distribution; the
      boxes show the inter-quartile range; the whiskers
      extend to either 1.5 times the inter-quartile range or the
      maximum or the minimum value of the data, whichever is larger; and the open
      circles show the outliers. The dashed lines show the magnitude
      limits of this catalog ($14 \lesssim r \leq 20$). The secondary
      distribution tracks the primary distribution at all distances.}
  \label{Fig: mass2}
  \end{centering}
\end{figure}

As discussed above, selection biases in {\slowpokes} are in general a
function of distance because of the large range of distances probed 
(see Sec.~\ref{Sec: pop}; Figures~\ref{Fig: sample_prop} and \ref{Fig:
  bin_grid}). Thus, in Figure~\ref{Fig: mass2} we show the
box-and-whisker plot of the distribution of the \rz\ colors of the
primary (blue) and secondary (red) components as a function of distance in 100~pc
bins. The bar inside the box is the median of the
distribution; the boxes show the inter-quartile range (defined as the
range between 25$^{\rm th}$ and 75$^{\rm th}$ percentiles); the
whiskers extend to either the 1.5 times the inter-quartile range or
the maximum or the minimum value of the data, whichever is larger; and the
open circles show the outliers of the distribution. The black dashed
lines show the bright and faint limits of our sample ($14 \lesssim r \leq 20$).
As the figure indicates the primary and secondary distributions are bluer
at increasing distances, as would be expected due to the bright and
faint limits; but the secondary distribution does not change as
compared to the primary distribution as a function of
distance. However, this is likely to be a strict function of the
faintness limit of our catalog: the stellar mass function peaks around
M4 spectral type \citepalias{Bochanski2010} while we cannot see M4 or
later stars beyond $\sim$400~pc. Hence, with our current sample, we cannot discern
whether the tendency for the secondary distribution to follow the primary
distribution is a tendency toward $q \sim 1$ or a tendency for the
secondary distribution to be drawn from the field mass function.

We can also examine the distribution of the secondary masses relative to
their primaries, i.e.\ the mass ratio distribution, which is an important
parameter in the study of binary star formation and evolution. As
shown in Figure~\ref{Fig: mass1} (right panel, solid histogram),
the distribution of mass ratios, $q \equiv M_2 / M_1$, is strongly
skewed toward equal-mass pairs: 20.9\%, 58.5\%, 85.5\% of pairs have
  masses within 10\%, 30\%, and 50\%, respectively, of each other. 
To determine whether this is strictly due to the magnitude limits in
\slowpokes, we calculated the mass ratios for hypothetical pairs with
the observed primary and the faintest observable secondary. The
resulting distribution (dashed histogram) is considerably different
from what is observed and shows that we could have identified pairs
with much lower mass ratios, within the $r \leq 20$ faintness
limit. The same result is obtained when we pair the 
observed secondaries with the brightest possible primary. Hence, 
we conclude that the observed distribution peaked toward
equal-masses among wide, low-mass stars is real and not a 
result of observational biases.

\subsection{Wide Binary Frequency}\label{Sec: binfrac}
\begin{figure}
  \begin{centering}
  \includegraphics[width=1\linewidth]{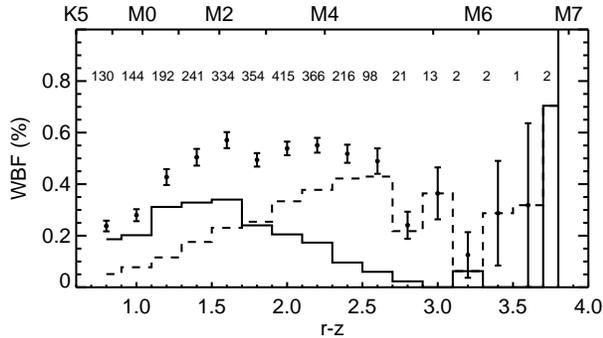}
    \caption{The wide binary fraction (WBF) of low-mass stars that are
      the primary (solid histogram) and secondary (dashed histogram)
      companions in wide pairs shown as a function of their \rz\
      colors, with spectral types shown along the top x-axis.
      The total WBF of low-mass stars in wide pairs , shown in black
      circles along with the binomial errors, shows a peak of $\sim$
      0.57\% for \rz\ $=$ 1.6 ($\sim$M2). The low WBF is expected due
      to the severe observational biases, as well as a reflection of
      our restrictive binary selection algorithm. The total number of
      CPM pairs in each bin are printed along the top.}
    \label{Fig: WBF1}
  \end{centering}
\end{figure}

To measure the frequency of wide binaries among low-mass stars, we
defined the wide binary frequency (WBF) as
\begin{equation}
{\rm WBF} = \frac{\rm number\ of\ CPM\ pairs\ found}{\rm number\ of\ stars\
  searched\ around}.
\end{equation}  
With 1342 CPM doubles among the 577459 stars that we searched around,
the raw WBF is 0.23\%. However, given the observational biases and our
restrictive selection criteria, 0.23\% is a lower
limit. Figure~\ref{Fig: WBF1} shows the WBF distribution as a function 
of \rz\ color with the primary (solid histogram), secondary (dashed
histogram), and total (solid dots with binomial error bars) WBF
plotted; the total number of pairs found in each bin are also shown
along the top. The WBF rises from $\sim$0.23\% at the bluest \rz\
color ($\sim$K7) to $\sim$0.57\% at \rz\ $=$ 1.6 ($\sim$M2), where it
plateaus. This trend is probably due at least in part to our better
sensitivity to companions around nearby early--mid M dwarfs as
compared to the more distant mid-K dwarfs or to the much fainter
mid--late M dwarfs. To get a first-order measure of the effects of the
observational biases, we can look at the WBF in specific distance
ranges where all stars of the given \rz\ colors can be seen and the
biases are similar for all colors, assuming the range of observable
magnitude are $r=$ 14--20. In particular, it would be useful to look
at \rz\ $=$ 0.7--1.5 where the WBF increases and \rz\ $=$ 1.5--2.5
where the WBF plateaus in Figure~\ref{Fig: WBF1}. Figure~\ref{Fig:
  WBF2} shows the WBF in the two regimes for $d=$ 247--1209~pc and
76--283~pc, respectively, where all companions of the given color range are
expected to be detected at those distances. In the restricted ranges,
the WBF is generally higher for a given color than in Figure~\ref{Fig:
  WBF1}, ranging between 0.44--1.1\%. More 
importantly, neither panel reproduces the trend in WBF seen for the same
color range in Figure~\ref{Fig: WBF1}, indicating that
observational biases and incompleteness play a significant role in the
observed WBF. As these two samples are pulled from two 
different distance ranges, some of the differences in the value of
WBF as well as the observed trend in WBF as a function of \rz\ might
be due to Galactic structure or age of the sampled pairs. Hence, even
the maximum observed WBF of 1.1\% in {\slowpokes} is likely
a lower limit on the true WBF.

\begin{figure}
  \begin{centering}
  \includegraphics[width=1\linewidth]{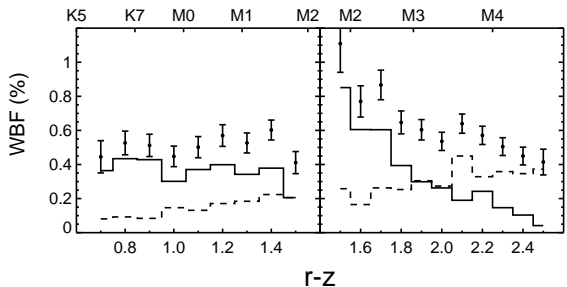}
    \caption{Same as Figure~\ref{Fig: WBF1}; but to explore the effect
      of biases in the WBF, we compare two 
      color ranges: \rz\ $=$ 0.7--1.5 where the WBF rises and \rz\ $=$
      1.5--2.6 where it plateaus in Figure~\ref{Fig: WBF1}. To make
      the selection effects similar throughout the selected color
      range, we restricted the distance of the stars such that all
      secondaries of the given color range can be seen throughout the
      distance range. The WBF seen here (i) is higher than in
      Figure~\ref{Fig: WBF1} for a given color bin and (ii) as a
      function of \rz\ does not reproduce the same trend as in
      Figure~\ref{Fig: WBF1}. Since the two samples were chosen from
      two different distance ranges, they might also reflect
      differences due to the age of the sampled pairs and/or the
      Galactic structure at that position. These differences indicate
      significant observational biases and incompleteness affect our WBF
      determination.}
    \label{Fig: WBF2}  
  \end{centering}
\end{figure}

\begin{figure}
  \begin{centering}
  \includegraphics[width=1\linewidth]{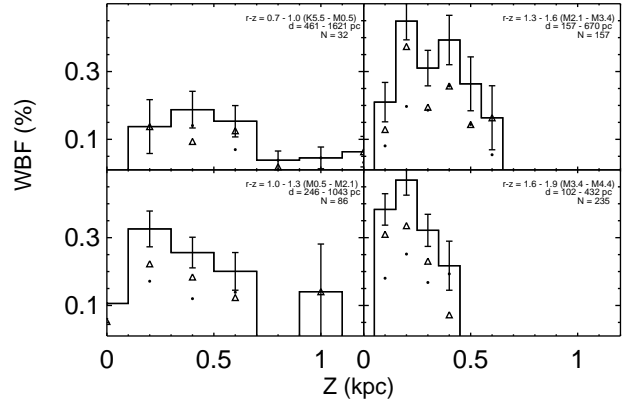}
    \caption{WBF, with binomial errors, decreases with
      Galactic height at all \rz\ color bins for the primaries
      (circle), secondaries (triangle), and the total (solid
      histogram). In order to keep biases similar in a color bin, we
      selected a distance range such that all stars with \rz\ colors
      within $\pm$0.3 dex of the bin could be seen throughout the distance
      range. This decrease in WBF with increasing $Z$ suggests
      dynamical destruction of (older) pairs at larger Galactic heights.}
    \label{Fig: WBF3}
  \end{centering}
\end{figure}

{\slowpokes} should be useful for studying the WBF as a function of
Galactic height ($Z$), which can be taken as a proxy for
age. Figure~\ref{Fig: WBF3} shows the WBF vs.\ $Z$, in \rz\ color
bins. Again, to keep the observational biases similar across color
bins, we selected distances ranges such that CPM companions with \rz\ colors
within $\pm$ 0.3 could be seen in SDSS; for example, in the first bin
(\rz\ $=$ 0.7--1.0) all companions with \rz\ $=$ 0.4--1.3 are expected
to be seen. Note that this approach is biased towards equal-mass pairs, with
high mass-ratio pairs never counted. Consequently, the WBF is lower
than in the Figures~\ref{Fig: WBF1} and \ref{Fig: WBF2}, with maximum
at $\sim$0.35\% for \rz\ $=$ 1.3--1.9; the WBF peaks at around the same
color range in all three figures. However, our method ensures that we
are sensitive to {\em all} similar mass pairs across all of the
distance bins. Figure~\ref{Fig: WBF3} suggests that the WBF decreases with
increasing Z.  As this trend appears for both primary and secondary
components for almost all spectral types (colors) probed, it is likely
not an artifact of observational biases and is strong evidence for the
time evolution of the WBF. Wide binaries are expected to be perturbed
by inhomogeneities in the Galactic potential, giant molecular
clouds, and other stars  and, as a result, to dissipate over time. As
pairs at larger Galactic heights are older as an ensemble, it is
expected that a larger fraction of pairs at larger $Z$, which are
older, have dissipated. 

\section{Discussion} \label{Sec: discussion}
As discussed above, \citetalias{Sesar2008} recently conducted a search
of CPM pairs in the SDSS DR and found $\sim$22000 wide pairs, of all
spectral types, with the probability of being a real binary of 
$\sim$67\% at angular separations of 5--30$\arcsec$. This search was
done for all stars in the DR6 photometry with $\mu\geq$15\masyr\ and
matched both components of proper motions within 5\masyr. As
\citetalias{Sesar2008} note, the sample is not very efficient for
follow-up studies due to the likely large number of chance alignments
(i.e.\ false positives). Our analysis in this paper indicates that
their low proper-motion cutoff is chief culprit for the contamination;
with a higher proper-motion cutoff similar to ours, the
\citetalias{Sesar2008} sample could in principle be sifted of many
false positives. \citetalias{Sesar2008} used a statistical subtraction
to correct the ensemble for the chance alignments, which suffices for
the purpose of their work. While {\slowpokes} is intended to serve a
very different purpose---namely to provide a ``pure" catalog of
high-fidelity systems suitable for targeted follow-up studies---it is
useful to compare some of our tentative findings and interpretations
with those of \citetalias{Sesar2008} based on their more complete
sample.

For example, the upper limit of the WBF in {\slowpokes} ($\sim$1.1\%; e.g.\ 
Figures~\ref{Fig: WBF2}) is consistent with the 0.9\% at $Z=$ 500~pc
from \citetalias{Sesar2008}. This suggests that, for specific color and distance ranges, 
the selection biases in {\slowpokes} may not significantly affect the
ability of this sample to characterize the WBF among late-type stars
at the widest separations. \citet{Kraus2009b} similarly find a WBF of
at most a few percent for masses 0.012--2.5 \Msun\ at separations of
500--5000~AU in their study of the Taurus and Upper Sco star-forming
regions. Moreover, they find that the distribution of binary
separations in their sample remains flat out to the limits of their 
survey ($\sim$5000~AU) for all but the lowest mass systems with
\Mtot $<$ 0.3\Msun. At the same time, all of these studies fall
well below the WBF (for $s>$ 1000~AU) of $\sim$9.5\% determined for
solar-type {\em Hipparcos} stars \citep{Lepine2007a} and the 10\%
suggested by \citet{Longhitano2010} for G5 or later stars in the solar
neighborhood from SDSS DR6. Similarly, integrating the 
Gaussian distribution of \citetalias{Duquennoy1991} suggests a WBF of 
$\sim$9.2\% for 1000~AU $\leq a \leq$ 0.5~pc and $\sim$0.64\% for $a>
$ 0.5~pc for solar-type stars. Even accounting for the lower multiplicity seen
in M dwarfs by \citetalias{Fischer1992}, this is much higher than our
result. Similarly, the decrease of WBF with increasing $Z$ in {\slowpokes}
(Figure~\ref{Fig: WBF3}) was noted by \citetalias{Sesar2008} in their
sample as well.

\citetalias{Sesar2008} had noted that the companions of
red stars, unlike the bluer stars, seemed to be drawn randomly from
the field stellar mass function. We do not find evidence to support
that finding, but our results are probably limited by the faintness
limit. With respect to the distribution of mass ratios 
(Figure~\ref{Fig: mass2}), \citet{Reid1997} observed, as we do, a
strong skew towards equal-mass pairs with a peak at $q\geq$0.8 in
their 8~pc sample. However, \citetalias{Fischer1992} found a flat
distribution in their sample of dMs within 20~pc. More recently,
\citet{Kraus2009b} also found that a bias towards equal-masses in their
study of 0.012--2.5~\Msun\ wide binaries (500--5000~AU separations) in the
Taurus and Upper-Sco star-forming regions. In contrast, solar-type
field stars exhibit a definitive peak at $q\sim$~0.3
(\citetalias{Duquennoy1991}, \citealt{Halbwachs2003}). With the field
mass function peaking around $\sim$M4 spectral type
\citepalias{Bochanski2010}, a peak at $q\sim$0.3 for solar-type stars
and at $q\sim$1 for M dwarfs are both consistent with the secondary
mass function being a subset of the field mass function.

With some systems wider than $\sim$1~pc, {\slowpokes} provides the
largest sample to date of the widest CPM doubles. However, these are not
the only systems known at this separation regime;  Table \ref{Tab:
  wide_binaries} lists 84 systems with projected physical separations
greater 0.05~pc ($\sim$10$^4$~AU). This sample is biased towards pairs
with at least one {\em Hipparcos} star, with about half of them from
the CPM catalog of \citet{Lepine2007a}. This is not a comprehensive list of
all known wide binaries; for example, the existing CPM catalogs of
\citet{Luyten1979a,Luyten1988}, \citet{Bahcall1981},
\citet{Halbwachs1986}, and \citet{Chaname2004} likely contain at least a few
hundred pairs in this separation range. Even among the brown dwarfs,
where most known pairs have separations smaller than 15~AU, two
very-wide systems---the 5100~AU 2MASS J0126555-502239
\citep{Artigau2007, Artigau2009} and the 6700~AU 2MASS J12583501+40130
\citep{Radigan2009}---have already been identified. All of these
wide pairs and other low-binding energy VLM systems from the
VLM binary archive \footnote{\url{http://vlmbinaries.org}} are shown,
when relevant, in Figures~\ref{Fig: bin_sep} and \ref{Fig: BE2}
for comparison with the {\slowpokes} distribution. Most of the CPM
doubles in Table~\ref{Tab: wide_binaries} were found in nearby, high
proper motion catalogs. Due to the depth of SDSS imaging and the
larger distances probed, {\slowpokes} significantly increases CPM pairs
with large physical separations. For the same reason, the existence of
pairs as wide as $\gtrsim$ 1~pc in {\slowpokes} (Figure~\ref{Fig:
  bin_sep}) should perhaps not be surprising.

\begin{figure}
  \begin{centering}
  \includegraphics[width=1\linewidth]{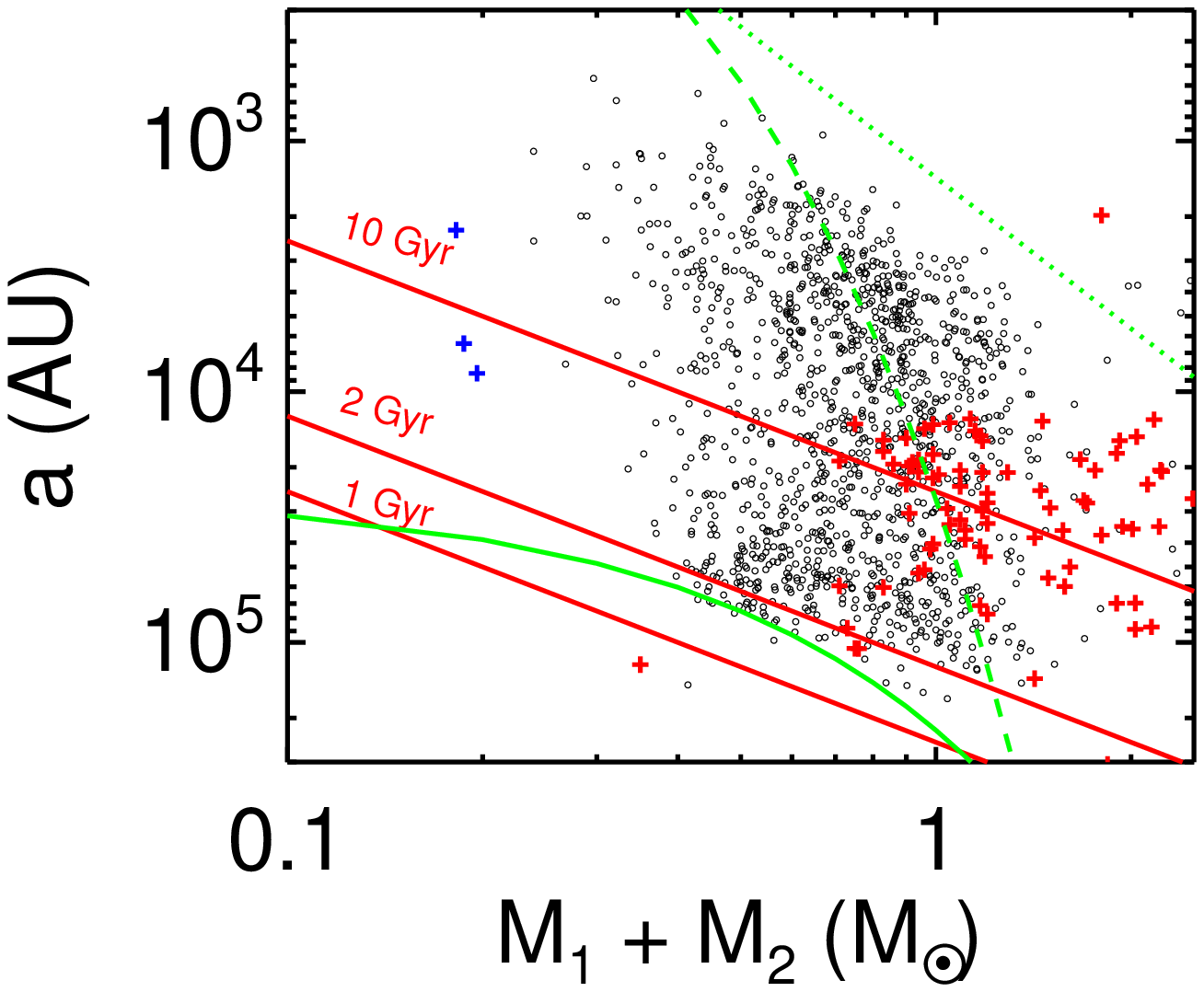}
  \caption{The physical separation of {\slowpokes} pairs (black
    circles) and previously known wide CPM doubles (red pluses;
    Table~\ref{Tab: wide_binaries}) and VLM binaries (blue pluses) as
    a function of the total 
    mass of the system, as inferred from their \rz\ colors. Previous
    studies by \citet{Reid2001b} (green dashed line) and
    \citet{Burgasser2003} (green dotted line) have also suggested
    empirical limits for stability. As neither of those describe the
    {\slowpokes} envelope, we have followed these previous authors and
    fit a log-normal (solid green line). Numerical simulations 
    by \citet{Weinberg1987} suggest that the lifetime of wide
    pairs is a function of age as well; the ``isochrones'' show the
    expected $a$ vs. \Mtot\ relationships for dissipation times of 1, 2, and 10~Gyrs
    (solid red lines), as defined by Eq.~(\ref{Eqn: stability}). While
    some very wide {\slowpokes} pairs are expected to dissipate on
    timescales of 1--2~Gyr, most should be stable for $\gtrsim$
    10~Gyrs. All SDs are excluded from this figure.}
  \label{Fig: BE1}
  \end{centering}
\end{figure}

\begin{figure*}
  \begin{centering}
  \includegraphics[width=0.7\linewidth]{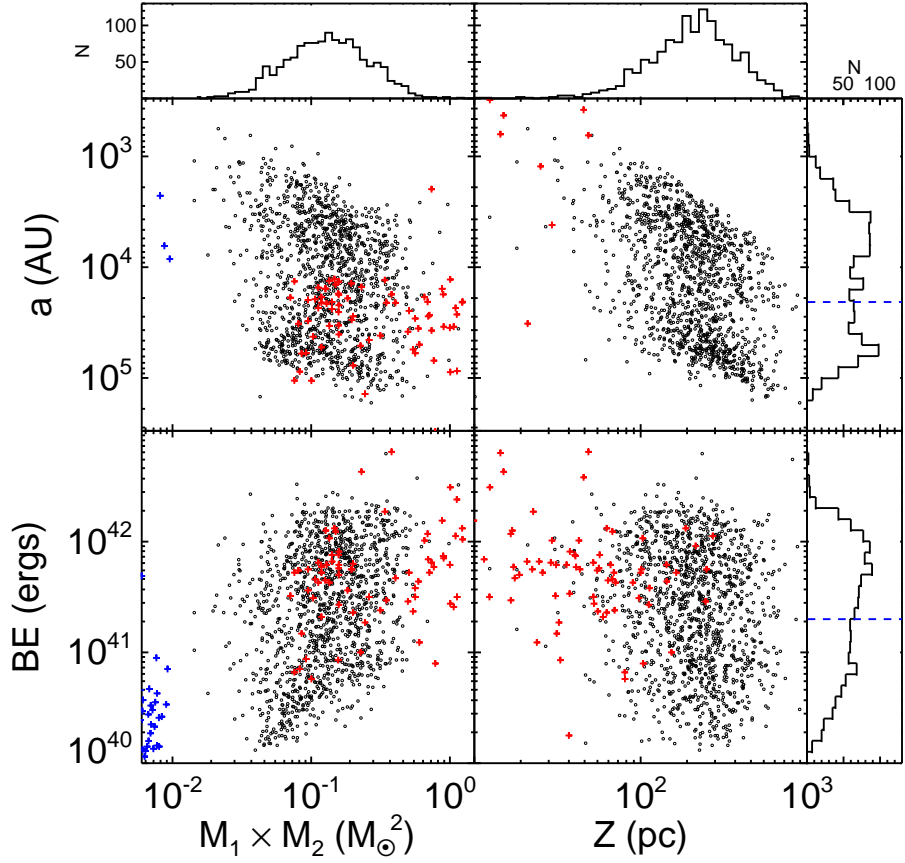}
    \caption{The physical separations and binding energies of
      {\slowpokes} pairs as a function of the product of their masses
      (left) or the Galactic height (right), with the histograms
      of the distributions shown along the top and the right. For
      comparison, other CPM pairs (red pluses; Table~\ref{Tab:
        wide_binaries}) and VLM binaries (blue pluses; VLM binary
      archive) from the literature are shown. The bimodality in
      physical separation (at $\sim$0.1~pc) seen in Figures~\ref{Fig:
        bin_sep} and \ref{Fig: BE1} and along the top panels in this figure is 
      evident in binding energy as well (at $\sim$10$^{41.3}$~ergs);
      these are marked with blue dashed lines in the histograms. As
      the binding energy is not dependent on $Z$, any selection
      effects or biases that depend directly or indirectly on $Z$ are
      not likely to be the cause of the bimodality. Previously
      suggested minimum binding energy for stellar ($10^{42.5}$~ergs)
      and substellar ($10^{40}$~ergs) binaries \citep{Close2003} are
      violated by most {\slowpokes} as well as other CPM and VLM
      pairs. All SDs are excluded from this figure.}
  \label{Fig: BE2}
  \end{centering}
\end{figure*}

At the same time, the question of how such wide, weakly bound systems survive
over time and how they form in the first place is a very interesting 
issue that the {\slowpokes} sample will be well suited to address.
The initial distribution of separations of wide binaries is not static
over time but is modified by interactions with other stars, molecular
clouds, and variations in the Galactic potential. Over the lifetime of
a binary, the small and dissipative, but numerous, encounters that it
undergoes with other stars is much more efficient at disrupting the
system than single catastrophic encounters, which are very rare
\citep{Weinberg1987}. Using the Fokker-Planck coefficients to describe
the advection and diffusion of the orbital binding energy due to such
small encounters over time, \citet{Weinberg1987} calculated that the
average lifetime of a binary is given by: 
\begin{equation}
  \begin{split}
    t_\star(a) \approx 18\ {\rm Gyr}
    \left(\frac{n_\star}{0.05\ {\rm pc}^{-3}}\right)^{-1}
    \left(\frac{M_\star}{\Msun}\right)^{-2} 
    \left(\frac{\Mtot}{\Msun}\right)\\
    \left(\frac{V_{\rm rel}}{20\ {\rm \kms}}\right)
    \left(\frac{a}{0.1 {\rm pc}}\right)^{-1} 
    (\ln \Lambda)^{-1}
  \end{split}
\end{equation}
where $n_\star$ and $M_\star$ are the number density and average mass of the
perturbers, $V_{\rm rel}$ is the relative velocity between the binary system
and the perturber, $\Mtot$ and $a$ are the total mass and semi-major
axis of the binary, and $\Lambda$ is the Coulomb logarithm. Using the
average Galactic disk mass density of 0.11\Msun~pc$^{-3}$, an average
perturber mass of 0.7\Msun, $V_{\rm rel}=$ 20\kms, and $\Lambda=$ 1
\citep[see also][]{Close2007}, we can rewrite the above equation as: 
\begin{equation}\label{Eqn: stability}
a \simeq 1.212\ \frac{\Mtot}{t_\star} {\rm pc}
\end{equation}
for $M$ in \Msun\ and $t_\star$ in Gyrs. This describes, statistically,
the widest binary that is surviving at a given age. For example, at
1~Gyr 1\Msun\ binaries as wide as 1.2~pc are likely to be bound, but
by 10~Gyr, all systems wider than 0.12~pc will likely have been
disrupted. Hence, the combination of Eq.~(\ref{Eqn: stability}) and
the distribution of binary separation at birth may describe the current
binary population of the Galaxy. Figure~\ref{Fig: BE1} shows the
physical separation vs.\ total mass for {\slowpokes} pairs and the
other known wide CPM pairs from the literature (see above);
characteristic disruption timescales of 1, 2, and 10 Gyr for
Eq.~(\ref{Eqn: stability}) are over-plotted. From this, most
{\slowpokes} pairs can be expected not to dissipate before 1--2~Gyr,
with approximately half of the population stable enough to last
longer than 10~Gyr. 

Interestingly, the widest {\slowpokes} systems appear to violate
previously proposed empirical limits on maximum separations
\citep[dashed and dotted green lines in Figure \ref{Fig: BE1};][]{Reid2001b,
  Burgasser2003}. Similarly, previously proposed empirical limits
based on binding energies, $10^{41}$~ergs for stellar and and
$10^{42.5}$~ergs very low-mass regimes \citep{Close2003, Close2007,
  Burgasser2007b}, are also violated by {\slowpokes} pairs as well as
other known CPM doubles. This is
clearly evident in Figure~\ref{Fig: BE2}, where we see that the most
weakly bound {\slowpokes} systems have binding energies of only $\sim
10^{40}$ ergs, comparable to the binding energy of Neptune about the
Sun\footnote{{\slowpokes} systems are much more likely than the
Sun--Neptune system to be disrupted by passing stars due to the
binaries' larger cross-section of interaction.}. \citet{Faherty2010}
have also noted such transgressions in their sample of wide, very
low-mass CPM doubles.

It seems that previously proposed empirical limits were too
restrictive as the widest and the VLM pairs that clearly violate them
were not available at the time. We can follow the approach of
\citet{Reid2001b} and \citet{Burgasser2003} to fit a log-normal that
forms an empirical envelope around the {\slowpokes} pairs
(Figure~\ref{Fig: BE1}, green solid lines). Although this new fit is
not likely to define the absolute stability limit for wide pairs, the
fit is notable for how well it describes the envelope. 

We note that Eq.~(\ref{Eqn: stability}) as well as most
previously proposed empirical limits have used the total mass of  
the system and do not take into account the mass ratio of the
components. This creates a degeneracy as the binding energy is
dependent on the product of masses. Instead, it may be more physical to
consider the wide, low-mass pairs as a function of the product of
the masses of their components---i.e., the binding energy (BE) of the system.
Figure~\ref{Fig: BE2} (left panels) present the $a$ and BE vs. $M_1
\times M_2$ for the {\slowpokes} pairs. That the distribution of $a$ as a
function of either sum or product of the component masses appears
qualitatively similar is a manifestation of the fact that {\slowpokes}
pairs tend to be equal mass (see Figure~\ref{Fig: mass1}); for high-$q$
pairs there would be a more noticeable difference in these distributions.

Perhaps, of particular interest to the issue of wide binary stability,
formation, and evolution is our observation of a distinctly bimodal
distribution of binary separations, seen in Figure~\ref{Fig: BE2}, which
equates to a bimodal distribution of system binding energies as well.
The break in the bimodal distribution at binding energy
$\sim$10$^{41.3}$~ergs is remarkable for how well it agrees with the
previously proposed empirical limit in binding energy for stellar
binaries to be dynamically stable \citep{Close2003, Close2007,
  Burgasser2007b}. In addition, as can be seen in Figure~\ref{Fig:
  BE1}, the 10~Gyr disruption 
``isochrone" of \citet{Weinberg1987} nicely traces the observed break in the
distribution over nearly a full decade of total system mass. The trend
of the break in the distribution with mass is also important because
it makes it very unlikely that the bimodality is the result of
observational biases in the {\slowpokes} sample. Similarly, there is no
effect of the heliocentric distances or the Galactic heights
(Figure~\ref{Fig: BE2}, right panels) of the pairs on the physical
separation or binding energy. As discussed earlier, the principal
observational biases in {\slowpokes} affect only (i) the smallest
separations most severely because of the strict cut-off at 7$\arcsec$
and (ii) to a lesser extent the largest separations as we limit our
search to 180$\arcsec$.

\begin{figure}
  \begin{centering}
  \includegraphics[width=0.8\linewidth]{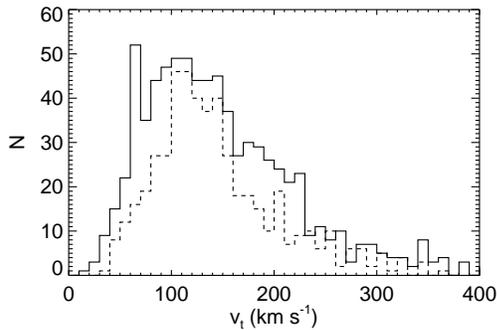}
    \caption{The tangential velocity ($v_t = 4.74~\mu~d$) distribution
      for the low-binding energy population (dashed
      histogram) and high-binding energy distribution
      (solid histogram), with the segregation at binding energy of
      $10^{41.3}$~ergs that was noticed in Figure \ref{Fig: BE2}. A smaller
      dispersion in the tangential velocity supports our suggestion
      that the low-binding energy distribution is of a younger
      population.}
    \label{Fig: vtan}
  \end{centering}
\end{figure}

Thus, the bimodal distribution of binary separations suggests two
populations of binaries, perhaps representing (i) systems of stars
that formed with sufficient binding energy to survive for the age of
the Galaxy and (ii) relatively young systems that formed within the
past 1--2~Gyr but that will likely not survive much
longer. Figure~\ref{Fig: vtan} shows that the dispersion in the
tangential velocity is indeed smaller for the ``young'' population,
relative to the ``old'' population that has had more time to get
kinematically heated.\footnote{One might expect lower
    mass stars to move with higher velocities relative to higher mass
    stars if equipartition of kinetic energy holds. However,
    observations have shown that the equipartition of kinetic 
    energy does not hold and that kinematic heating is purely a
    function of age for field stars. Unlike in clusters,
    individual stars can be thought of as small, point masses when
    compared to the much larger Galactic potential or the giant
    molecular clouds that cause dynamical heating
    \citep{Bochanski2007a, Schmidt2010}.}
Quantitatively, the median absolute deviations
for the two populations are 46.1~\kms\ and 32.7~\kms. Curiously,
the separations of the binaries in the second group (up to 1~pc;
Figure~\ref{Fig: bin_sep}, right panel) is larger than the typical
sizes of prestellar cores ($\sim$0.35~pc), suggesting that these 
systems may not have formed via the ``standard" collapse and
fragmentation of individual cloud cores.

Recent N-body simulations predict bimodal distributions of
wide binary separations and may provide important theoretical insight
for understanding the formation and evolution of wide pairs. Although
the specific simulations predict bimodality on different scales than
seen in our observations, some of the resulting predicted physical
properties of pairs are similar to those found in {\slowpokes}.
For example, \citet{Kouwenhoven2010} 
find that the distribution of binary semi-major axes will in general
be bimodal on the size scale of star-forming clusters (i.e.\ $\sim
0.1$~pc), with a group of tight systems formed primarily via dynamical
interactions near the cluster center and a second group of loosely
bound systems formed through random pairing during the slow
dissolution of the cluster. In their simulations, up to $\sim 30$\% of
the resulting binaries will be in the latter ``dissolution peak", with
separations of $\sim 0.1$~pc. They predict that the mass ratios of
these systems will reflect random pairing. Alternatively,
\citet{Jiang2009} have found that a bimodal distribution arises due to
the slow (as opposed to instantaneous), diffusive disruption of
systems on timescales of several Gyrs, and that the details of the
resulting bimodality is mostly independent of the initial distribution
of separations arising from the formation process. Their
simulations predict that the bimodal break in the distribution occurs
at a few Jacobi radii, which corresponds to a separation of $\sim
2$~pc for a total system of 1\Msun. Perhaps the diffusive dissipation
mechanism is more efficient than predicted by \citet{Jiang2009}, or
perhaps the ``tidal-tail peak" of dissolving systems has not yet been
observed. In the future, the {\slowpokes} catalog can be extended to
even larger physical separations to probe this possibility. 

\section{Conclusions}\label{Sec: conclusions}
We have created the {\slowpokes} catalog, comprising 1342 CPM binary
pairs, identified through statistical matching of 
angular separation, photometric distances, and proper motion
components. We have sifted the sample of chance alignments using a
Galactic model based on empirical observations of stellar spatial and
kinematic distributions. With the objective that each pair can be
confidently used to investigate various science questions regarding
low-mass stars, we have adopted a very restrictive set of selection
criteria. This approach clearly underestimates the number of binary
systems. Moreover, the sample includes several biases, the most
important of which are a lack of systems with physical separations
smaller than $\sim 1000$~AU (arising from a strict bias against
angular separations smaller than 7$\arcsec$), and the exclusion of
certain types of higher-order multiples (e.g.\ triples) due to the
strict photometric distance matching. However, as a consequence the
catalog should contain very few false positives, making follow-up
studies efficient.  

We built a Monte Carlo-based six-dimensional Galactic model that is able to
replicate the positional and kinematic properties of the stars in the
Milky Way. In its current incarnation, we used it to calculate the
the number of stars within a certain spatial volume in the Galaxy and
the likelihood that those stars have common kinematics (proper
motions) by chance. One of the things this model underscores is how
difficult it is to find two physically unassociated stars close
together in space: along a typical SDSS LOS, there are
expected to be only 0.52 chance alignments within 15$\arcsec$ and a
minuscule 0.03 chance alignments if the volume is considered. The
additional matching of proper motions gives each of the accepted
{\slowpokes} binaries a very low probability of being a false positive.

Due to their intrinsic faintness and the resulting small numbers,
binarity studies of low-mass stars have been limited in
scope. However, with the advent of large-scale deep surveys, detailed
and statistically significant studies of M (and L) dwarfs are being
done. {\slowpokes} is now the largest sample to date of very wide,
low-mass binaries. In particular, {\slowpokes} provides a large sample
of systems with physical separations up to $a\sim$ 1~pc that will be
useful for putting firm constraints on the maximum size of physically
associated systems. How the widest of these systems form, and how long
they survive, is in particular an interesting question that
{\slowpokes} is well suited to address. While numerical calculations
suggest that approximately half of {\slowpokes} systems can remain
bound for at least 10~Gyr \citep{Weinberg1987}, previously proposed
empirical limits are violated by many {\slowpokes} systems.

Indeed, the distribution of {\slowpokes} binary separations is
distinctly bimodal, suggesting the presence of (i) a population of
tightly bound systems formed with sufficient binding energy to remain
intact for the age of the Galaxy and (ii) a population of weakly bound
systems that recently formed and that are unlikely to survive past
1--2~Gyr. Recent N-body simulations \citep[e.g.][]{Kouwenhoven2010,
  Jiang2009} in fact predict a bimodal distribution of binary
separations on the scales probed by the widest {\slowpokes} systems. 

We observed a wide binary frequency of $\sim$1.1\%, which is likely
to be a minimum given the nature of our sample. While this is
consistent with the results from \citetalias{Sesar2008} who found 0.9\% of
stars at $Z=$ 500~pc had wide companions., it is significantly lower
than $\sim$9.5\% of nearby solar-type {\em Hipparcos} stars having wide
companions \citep{Lepine2007a}. While the incompleteness involved in
the searching for companions at large distances probably causes some
of this, both this study and \citetalias{Sesar2008} saw a decrease in
binary fraction as a function of Galactic height, evidence of
dynamical destruction of older systems. Hence, the wide binary
fraction around our initial sample of low-mass stars might actually be
significantly lower than the {\em Hipparcos} stars.

Besides the importance of {\slowpokes} for constraining
models of formation and evolution of binary stars, {\slowpokes}
systems are coeval laboratories---sharing an identical formation and
evolutionary history without affecting each other---making them
ideal for measuring and calibrating empirical relationships between
rotation, activity, metallicity, age, etc. We have started programs to
test and calibrate the age--activity relationship measured by
\citet{West2008} and to explore whether gyrochronology
\citep{Barnes2003a, Barnes2007} can be applied in the fully-convective
regime. As coeval laboratories allow for the removal of one or more of
the three fundamental parameters (mass, age, and metallicity), much
more science can be done with a large sample of such systems.

Future astrometric missions, such as the Space Interferometry Mission
(SIM), should provide exquisite astrometry, perhaps enabling us to
trace the orbits of some of the {\slowpokes} systems. While tracing
orbits with periods $\gtrsim10^{4-6}$~years sounds ambitious, with
SIM's microarcsecond level (or better) astrometry \citep{Unwin2008}
combined with SDSS, DSS, and/or other epochs, it is not
unrealistic. Similarly, the ``identical'' twins in {\slowpokes} would
be ideal sites to probe for the presence and differences in the
formation mechanism of planets. As each identical twin in a CPM double
provides similar environment for the formation and evolution of planets,
these systems can be ideal sites to study planetary statistics. Due to
their large separations the stars are not expected to influence each others
evolution but have similar mass, age, and metallicity, as we noted earlier in 
\S~\ref{Sec: intro}.

The {\slowpokes} catalog, as the name suggests, only contains
systems for which kinematic information is available. We can, however,
use the results from the Galactic model to identify pairs at the small
separations ($\dtheta<7\arcsec$), albeit with a larger uncertainty,
for which no kinematic information is available. Similarly companions
which are fainter than $r=$ 20 can also be identified as the SDSS
photometry is complete to $r=$ 22.5. The latter systems are likely to
be skewed towards late-type dMs and unequal-mass pairs. A follow-up
paper will study such systems and will add a large proportion of wide
systems.

\section*{Acknowledgments}
We thank the anonymous referee for insightful and useful
comments. We would also like to thank Gibor Basri, Adam J. Burgasser, Julio
Chanam{\'e}, Kevin R. Covey, Paul Harding, Kelly Holley-Bockelmann,
Hugh C. Harris, Eric L. N. Jensen, Suzanne L. Hawley, Nicholas M. Law,
and Heather L. Morrison for discussions and/or feedback at various points in
the project. SD was funded by NSF grant AST-0909463 (K. Stassun,
PI). AAW and JJB thank Adam J. Burgasser for financial 
support. This work was sponsored in part by the National Aeronautics
and Space Administration, as part of the Space Interferometry Mission
Science Studies, through a contract with the Jet Propulsion
Laboratory, California Institute of Technology.

Funding for the SDSS and SDSS-II has been provided by the Alfred
P. Sloan Foundation, the Participating Institutions, the National
Science Foundation, the US Department of Energy, the National
Aeronautics and Space Administration, the Japanese Monbukagakusho, the
Max Planck Society, and the Higher Education Funding Council for
England. The SDSS Web Site is http://www.sdss.org. 

The SDSS is managed by the Astrophysical Research Consortium for the
Participating Institutions. The Participating Institutions are the
American Museum of Natural History, Astrophysical Institute Potsdam,
University of Basel, University of Cambridge, Case Western Reserve
University, University of Chicago, Drexel University, Fermilab, the
Institute for Advanced Study, the Japan Participation Group, Johns
Hopkins University, the Joint Institute for Nuclear Astrophysics, the
Kavli Institute for Particle Astrophysics and Cosmology, the Korean
Scientist Group, the Chinese Academy of Sciences (LAMOST), Los Alamos
National Laboratory, the Max-Planck-Institute for Astronomy (MPIA),
the Max-Planck-Institute for Astrophysics (MPA), New Mexico State
University, Ohio State University, University of Pittsburgh,
University of Portsmouth, Princeton University, the United States
Naval Observatory, and the University of Washington.

We acknowledge use of the SIMBAD database, maintained
by Strasbourg Observatory; the VLM Binaries Archive, maintained by Nick
Siegler, Chris Gelino, and Adam Burgasser; and the ADS
bibliographic service.

\bibliography{ads,inprep}
\end{document}